\documentclass[12pt,preprint,amssymb]{aastex}

\usepackage{graphicx}

\usepackage[figuresright]{rotating}

\def\suzaku{{\em Suzaku}\/}

\def\chandra{{\em Chandra}\/}
\def\asca{{\em ASCA}\/}

\def\xmm{{\em XMM-Newton}\/}

\def\arcmin{$^{\prime}$}

\shorttitle{suzaku observation of A1795}
\shortauthors{Gu et al.}

\begin{document}
\title{Two-Phase ICM in the Central Region of the Rich Cluster of
  Galaxies Abell 1795: A Joint {\em Chandra}, {\em XMM-Newton}, and
  {\em Suzaku} View}

\author {Liyi Gu\altaffilmark{1,2}, Haiguang Xu\altaffilmark{2},
  Junhua Gu\altaffilmark{2}, Madoka Kawaharada\altaffilmark{3}, 
  Kazuhiro Nakazawa\altaffilmark{1}, Zhenzhen
  Qin\altaffilmark{2}, Jingying
  Wang\altaffilmark{2}, Yu Wang\altaffilmark{4}, 
  Zhongli Zhang\altaffilmark{5}, Kazuo
  Makishima\altaffilmark{1}}

\altaffiltext{1}{Department of Physics, University of Tokyo, 7-3-1
  Hongo, Bunkyo-ku, Tokyo 113-0011, Japan}
\altaffiltext{2}{Department of Physics, Shanghai Jiao Tong University,
800 Dongchuan Road, Minhang, Shanghai 200240, PRC}
\altaffiltext{3}{Institute of Space and Astronautical Science (ISAS),
  Japan Aerospace Exploration Agency (JAXA), 3-1-1 Yoshinodai,
  Chuo-ku, Sagamihara, Kanagawa 252-5210, Japan}
\altaffiltext{4}{Shanghai Astronomical Observatory, Chinese Academy of
  Sciences, Nandan Road 80, Shanghai 200030, PRC}
\altaffiltext{5}{Max-Planck Institut f\"{u}r Astrophysik, Karl-Schwarzschild-Stra\ss e 1, 
D-85740 Garching, Germany}

\begin{abstract}

  Based on a detailed analysis of the high-quality {\it Chandra}, {\it
    XMM-Newton}, and {\it Suzaku} data of the X-ray bright cluster of
  galaxies Abell 1795, we report clear evidence for a two-phase
  intracluster medium (ICM) structure, which consists of a cool (with
  a temperature $T_{\rm
    c}\approx 2.0-2.2$ keV) and a hot ($T_{\rm h}\approx 5.0-5.7$ keV)
  component that coexist and dominate the X-ray emission at least in the
  central 80 kpc. A third weak emission component
  ($T_{\rm 3}\approx 0.8 $ keV) is also detected within the
    innermost 144 kpc
  and is ascribed to a portion of inter-stellar medium (ISM) of the cD
  galaxy. Deprojected spectral analysis reveals flat radial
  temperature distributions for both the hot phase and cool phase components. These
  results are consistent with the \asca\ 
  measurements reported in Xu et al. (1998), and resemble the
  previous findings for the Centaurus cluster (e.g., Takahashi et
  al. 2009). By analyzing the emission measure ratio and gas metal
  abundance maps created from the {\it Chandra} data, we find that the
  cool phase component is more metal-enriched than the hot phase one
  in $50-100$ kpc region, which agrees with that found in M87 (Simionescu et
  al. 2008). The coexistence of the cool phase and hot phase ICM cannot be
  realized by bubble uplifting from active galactic nuclei (AGN)
  alone. Instead, the two-phase ICM
  properties are better reconciled with a cD
  corona model (Makishima et al. 2001). In this model, the cool phase
  may be ascribed to the plasmas confined in 
  magnetic loops, which are surrounded by the intruding hot phase ICM and have
  been polluted by metals synthesized in the
  cD galaxy. AGN feedback energy released in the
  innermost 10 kpc can serve as the heating source to
  prevent the loop-interior gas from cooling down to temperatures much
  lower than the observed value. The total gravitating mass profile
  exhibits a hierarchical structure, regardless of the ICM temperature
  modeling.

\end{abstract}
\section{INTRODUCTION}

First revealed by the {\em Einstein} observatory, the spectroscopic
temperature of the intracluster medium (ICM)
in many cD clusters of galaxies decreases toward the center by a factor of
$2-3$ (e.g., Canizares et al. 1988; Mushotzky \& Szymkowiak 1988; see also Fabian 1994
for a review). Generally, the appearance of such a central cool
component (hereafter CCC) has been described in two
possible but competing ways. One is a single-phase (hereafter 1P) scenario, in which
a mono-nature plasma showing an inward temperature decrease is assumed to permeate the whole
cluster (e.g., Allen et al. 2001; Vikhlinin et al. 2006). The other is a two-phase (hereafter 2P)
scenario, in which the cluster's central region is assumed to be
occupied by a mixture of two gas components characterized by
discrete temperatures (i.e., hot and cool), of which the relative
volume filling factors slowly vary with radius (e.g.,
Fukazawa et al. 1994, 1998; Takahashi et al. 2009; see Makishima et al. 2001 for a review). Although
in many cases it is difficult to distinguish between the 1P and 2P models due to insufficient
data quality, the implications of these models on the formation
and evolution of the CCC are intrinsically 
different. In the 1P scenario the CCC is simply interpreted as the dense cluster core, where radiative
losses have significantly lowered the gas temperature (see, e.g.,
Peterson \& Fabian 2006 for a review), while 
the 2P scenario favors an interpretation that the CCC is caused
  by the cD galaxy, rather than being a cooling portion
of the ICM (Makishima et al. 2001 and references therein). Furthermore, the choice
between the two scenarios has considerable impact on the accuracies of the measurements
of both gas metallicities (e.g., the ``Fe-bias'' discussed in Buote
2000) and, in some extreme cases, gravitating mass distributions
in clusters. For the latter one, the systematic mass bias induced by different ICM models can reach
$\sim 10$\% for the cluster's central region,
which is comparable to the typical deviation between the results
obtained in X-ray and lensing studies
(e.g., Nagai et al. 2007; Mahdavi et al. 2008). Hence, it is of fundamental and urgent importance
to determine which of the two scenarios is valid, based on the analysis of existing
high quality X-ray data of nearby, bright clusters.

So far, only a few works have been done to examine the preference
  between the 1P and 2P ICM models of
cD groups and clusters. Of these, Fukazawa et al. (1994) and Ikebe et al.
(1999) used the {\em ASCA} data to show that a 2P model with temperatures of 1.4 keV and 4.4 keV for the two components is required to describe the CCC of
the Centaurus cluster. Their results were recently confirmed by the analysis of the high quality data acquired
with the European Photon Imaging Camera (EPIC) and Reflection Grating Spectrometer (RGS) onboard \xmm\ (Takahashi
et al. 2009). The preference for the 2P scenario was also reported in the works on the Fornax cluster
(Ikebe et al. 1996; Matsushita et al. 2007) and the NGC 5044 group (Buote \& Fabian 1998; Buote et al. 2003;
Tamura et al. 2003) with {\em ASCA}, {\em Chandra}, {\em XMM-Newton},
and {\em Suzaku} observations. However, due to the lack of such studies on more representative relaxed clusters
or groups that exhibit strong CCCs, it is not yet clear whether or not the 2P scenario do surpass
the 1P scenario in general, and if yes, to what extent the 2P model can be applied to constrain the origin of
the CCC.

To address these issues, we study in this work the rich, bright cD
cluster of galaxies Abell 1795 (A1795 hereafter), by jointly analyzing
the archival {\em Chandra}, {\em XMM-Newton}, and \suzaku\ data. A1795 is suited to our
purpose as it is a relaxed, nearby (at a redshift of z $=0.0625$) cluster of galaxies
with a luminous X-ray emission
(with a $2.0-10.0$ keV luminosity of $L_{X}\approx 1.0\times10^{45}$ ergs $\rm s^{-1}$; Xu et al. 1998), which is found
peaked at the giant cD galaxy PGC 049005 (Jones \& Forman 1984; Briel \&
Henry 1996). Under the 2P assumption,
Xu et al. (1998) resolved a cool ($T_{X} \approx 1.7$ keV) and a hot
($T_{X} \approx 6.5$ keV) component in
the central 220$h_{71}^{-1}$ kpc region of this cluster with {\em
  ASCA}, although Ettori et al. (2002) employed the
1P model instead to fit {\em Chandra} spectra extracted from
the same region, and found an inward temperature decrease
to $\approx 2.5$ keV. In the past 11 years, X-ray observations of this
cluster have been accumulated to $\approx 300$ ks, 140 ks, and 270 ks
with {\it Chandra}, {\it XMM-Newton}, and {\it Suzaku}, respectively. Clearly this
is an ideal target for utilizing as much of existing high-quality
X-ray data as possible to judge which spectral model is more correct.



The layout of this paper is as follows. Section 2 gives a brief description
of our data reduction
procedure. The data analysis and results are described in \S 3. We
discuss the physical implication of our analysis in \S 4, and
summarize our work in \S 5. Throughout the
paper we assume a Hubble constant of $H_0=71$$h_{71}$ km s$^{-1}$
Mpc$^{-1}$, a flat universe with the cosmological parameters of
$\Omega_M=0.27$ and $\Omega_\Lambda=0.73$, and quote errors by the
90\% confidence level unless stated otherwise. At the redshift of this
cluster, 1\arcmin\ corresponds to about 72$h_{71}^{-1}$ kpc. To compare with previous
results, we adopt the solar abundance standards of Anders \& Grevesse (1989).

\section{OBSERVATION AND DATA REDUCTION}

\subsection{X-ray Observation}

\subsubsection{{\it Chandra}}

In Table 1, we list 11 \chandra\ datasets of A1795 obtained with
the advanced CCD imaging spectrometer (ACIS),
which are all used in the analysis. Among the 11 observations, five
ACIS-S3 exposures focused on the cluster center, and six ACIS-I
exposures covered off-center regions up to $r \sim 1.5$ Mpc. All data 
were telemetered in VFAINT mode, except for the one with ObsID 494 in FAINT 
mode. Using CIAO v4.4 and CALDB v4.4.7, we removed bad pixels and 
columns, as well as events with \asca\ grades 1, 5, and 7. We then
executed the gain, CTI, and astrometry corrections. The cumulative exposure
time was reduced from $\approx 170$ ks to $\approx 163$ ks after removing
intervals contaminated by occasional background flares with count rate
$>$$20$\% of mean value, which were detected by
examining $0.3-10.0$ keV lightcurves extracted from source free
regions near the CCD edges (e.g., Gu et al. 2009). When available, ACIS-S1
data were also used to crosscheck the determination of the
contaminated intervals. The obtained clean exposure time for each observation is listed
in Table 1. In Figure 1{\it a} we show the combined ACIS
image, which has been corrected for exposure but not for background.

\subsubsection{{\it XMM-Newton}}

\xmm\ was used to observe the central and the south periphery
regions of A1795 on 2000 June 26
and 2003 January 13, respectively, both with the EPIC operated in the full window mode with
thin filter, and with the RGS in the
spectroscopy mode (Table 1). Data reduction and calibration were carried out
with SAS v11.0.1. In the screening process we set $FLAG=0$,
and kept events with $PATTERNs$ 0--12 for MOS cameras and events with $PATTERNs$ 0--4
for pn camera. By examining lightcurves extracted in
$10.0-14.0$ keV and $1.0-5.0$ keV from source free regions, we rejected time intervals
affected by hard- and soft-band flares, respectively, in
which the count rate exceeds a 2$\sigma$ limit above the quiescent
mean value (e.g., Katayama et al. 2004; Nevalainen et al. 2005).
The cleaned MOS1, MOS2, and pn datasets of the central pointing have exposure times of 43.6
ks, 40.3 ks, and 30.2 ks, respectively (Table
1). The RGS data were screened using the method described in Tamura et al. (2001).

\subsubsection{{\it Suzaku}}

As listed in Table 1, A1795 was also observed by \suzaku\ on 2005
  December 10, with five pointings aimed at its 
central region, near south and north regions (offset $\approx 12^{\prime}$), and
far south and north regions (offset $\approx 24^{\prime}$). The onboard
X-ray Imaging Spectrometer (XIS; Koyama et al. 2007)
and the Hard X-ray Detector (HXD; Takahashi et al. 2007) were both operated in
normal modes. The same XIS datasets were already utilized by
  Bautz et al. (2009). In analyzing the obtained data, we used the software
HEASoft 6.11.1 and latest CALDB (20111109 for the XIS and 20110913 for
the HXD). We started with version 2.0
processing data, and removed the data obtained either near South Atlantic
Anomaly or at low elevation angles from the Earth rim
($<5^{\circ}$ and $<20^{\circ}$ for night and day,
respectively). Cut-off rigidity criteria of $> 8$ GV for HXD-PIN data
were also applied. For the XIS instrument, we further examined $0.3-10.0$ keV
lightcurves of a source free region in each CCD, and
filtered off anomalous time bins with count rate above the 2$\sigma$
limit of the quiescent mean value. The combined XIS image is shown in Figure 1{\it b}.
The data obtained in the far north field are contaminated seriously by solar wind charge exchange
emission (Fujimoto et al. 2007), and thus are not included in the
following data analysis. The obtained clean exposure times are listed
in Table 1. We do not consider HXD-GSO data in this paper.

\subsection{Point Sources, Background, and Systematic Uncertainties}

We excluded point sources detected beyond a $3\sigma$ threshold on the ACIS images
with the CIAO tool {\tt celldetect}, and masked the corresponding regions on the
EPIC and XIS images after considering the differences in the Point Spread Functions (PSFs) of
the instruments. The mask regions have radii of $12^{\prime\prime}$
and $70^{\prime\prime}$ for the EPIC and XIS, respectively. The point sources outside the ACIS sky
coverage were detected and masked using the EPIC images. The ancillary
response files (ARFs) and the redistribution matrix
files (RMFs) were calculated with the CIAO tools {\tt mkwarf} and {\tt
  mkacisrmf} for the ACIS, and
with the SAS tools {\tt arfgen} and {\tt rmfgen} for the EPIC. We calculated the RMFs for the RGS
using the SAS tool {\tt rgsrmfgen}, after blurring the line spread function by convolving the
raw RMFs with the surface brightness profile (\S 3.3.1) extracted from
the $0.5-8.0$ keV ACIS image along the dispersion direction. As for the XIS, we
generated the response files using {\tt xissimarfgen} (version 2008-04-05) and
{\tt xisrmfgen} (version 2007-05-14). To enhance the precision of the
Monte-Carlo simulation carried out by {\tt xissimarfgen}, we created the initial
feed photon list by combining all the exposure-corrected and background-subtracted
ACIS-I images. The method to
build the response of HXD-PIN is described in \S 3.4.

We estimated the background as a combination of three independent
components, i.e., non X-ray background (NXB), cosmic X-ray background
(CXB), and Galactic emission. For all the XIS data, we created the NXB spectra
using a dark earth observation database. The CXB and
Galactic emission components in the XIS data were estimated by analyzing a spectrum extracted from
a region about $26^{\prime}-30$\arcmin\ ($\approx
1.9-2.2$$h_{71}^{-1}$ Mpc at the distance of
A1795) south of the cluster center, which is covered by the dataset
with ObsID = 800012050 (Table 1); a similar background region was used in Bautz et al. (2009). 
Since this region is outside the virial
radius $r_{\rm 200} = 1.9$$h_{71}^{-1}$ Mpc, where the brightness of cluster emission is expected to be $\sim 10^{-13}$ $\rm ergs$ $\rm cm^{-2}$ $\rm
s^{-1}$ $\rm deg^{-2}$ in $0.5-2.0$ keV (Roncarelli et al. 2006),
lower than the detection limit of the XIS, i.e., $\sim 10^{-12}$ $\rm ergs$ $\rm cm^{-2}$ $\rm
s^{-1}$ $\rm deg^{-2}$ (Bautz et al. 2009), the ICM component can be ignored in the background fitting. After
subtracting the NXB from the extracted spectrum, we fitted 
the resulting spectrum with an absorbed power law model (photon index
$\Gamma = 1.4$) describing the CXB, and two unabsorbed optically thin thermal models
(abundance = 1 $Z_\odot$, temperatures = 0.08 keV 
and 0.2 keV; Snowden et al. 1998) describing the Galactic emission. The $2.0-10.0$ keV CXB flux was
estimated as $6.6\times10^{-8}$ $\rm ergs$ $\rm cm^{-2}$ $\rm s^{-1}$
$\rm sr^{-1}$, which is consistent with the results of Kushino et
al. (2002). The obtained CXB + Galactic
background template was applied to all the subsequent
XIS spectral analysis assuming the same uniform CXB + Galactic emission distribution over
the field of view (e.g., Sato et al. 2008). 

In order to achieve an accordant background model for the {\em Chandra} ACIS, we utilized
the above CXB + Galactic background template and calculated the ACIS
NXB spectra in an adaptive way as follows. For all ACIS-I
observations, background spectra were
extracted from regions $17^{\prime}-21$\arcmin\ ($\sim 1.3-1.5$$h_{71}^{-1}$ Mpc) away
from the cluster center, where the ICM brightness was measured to be
weak but cannot be neglected, on a level of 10\% to 40\% of the CXB + Galactic
components (Bautz et al. 2009). We subtracted the CXB + Galactic
components from the extracted spectra, whose brightness was fixed at the
XIS result obtained above, and fitted the resulting spectra
with an empirical NXB model that consists of a broken power law 
and five narrow Gaussian lines (e.g., Humphrey \& Buote 2006), plus a thermal APEC
model for the cluster emission. Since the
temperature and metal abundance of this ICM component cannot be well
constrained, due to the relatively high
NXB level of the ACIS, we fixed them to previous {\em Suzaku} results (temperature =
2.5 keV, abundance = 0.1 $Z_\odot$) reported in Bautz et
al. (2009). In the $r=17^{\prime}-21$\arcmin\ region, average surface brightness of the ICM
component was obtained as $1.7\pm0.8 \times 10^{-12}$ $\rm ergs$ $\rm cm^{-2}$ $\rm
s^{-1}$ $\rm deg^{-2}$ in $0.5-2.0$ keV, which is about $25$\% of the
CXB + Galactic components, in rough consistent with Bautz et al. (2009).
The ACIS-I NXB model was thus determined. The same procedure was applied to all ACIS-S data, but
source free regions on the S1 chip were used as the background regions
instead. To crosscheck above background model with the blank sky
templates, we applied both backgrounds to the spectral analysis in
\S3.1.1. By fitting the background-subtracted ACIS spectra extracted from
$9^{\prime}.8-17^{\prime}.0$ with a single-phase thermal model, we
found that the ICM temperatures with the two background models differ
by 0.35 keV, which is less than the 1$\sigma$ uncertainties of 0.5 keV. The
error caused by background model is much smaller for the inner
regions, where the ICM emission contributes to
$>80$\% of the total counts in $0.5-10.0$ keV (Table 2).

The background model for the {\em XMM-Newton} EPIC data was determined
with a method similar to that used for the {\em Chandra} ACIS
data. The EPIC spectra extracted from $19-25$\arcmin\ ($\sim
1.4-1.8$$h_{71}^{-1}$ Mpc) away from
the cluster center, covered by the data with ObsID = 0109070201, were
fit with a model consisting of the CXB + Galactic components, the EPIC NXB model 
including a broken power law and six Gaussian lines (e.g., Gastaldello et al. 2007; Zhang et al. 2009),
and a cluster component approximated by an APEC model. The temperature and
abundance of the cluster component were fixed at 2.4 keV and 0.1
$Z_\odot$ (Bautz et al. 2009), respectively, and the $0.5-2.0$ keV surface
brightness was obtained as $1.4\pm1.1 \times 10^{-12}$ $\rm ergs$ $\rm cm^{-2}$ $\rm
s^{-1}$ $\rm deg^{-2}$ ($\approx 15$\% of the CXB + Galactic components), in
agreement with Bautz et al. (2009) and our 
{\em Chandra} estimate.
The best-fit CXB + NXB + Galactic models were used to create background spectra.
In Figure 2{\it a}, we plot the NXB-subtracted background spectra
  against the best-fit CXB and Galactic components for the ACIS, EPIC,
  and XIS, as well as the cluster emission component for the former two instruments. For the RGS data, we employed the blank sky template (e.g.,
Takahashi et al. 2009) in the following spectral analysis.


Errors quoted below in the spectral fittings were estimated by
taking into account both statistical and
systematic uncertainties. The former was calculated by scanning
over the parameter space with
the {\tt XSPEC} command {\em steppar}, as the fitting was repeated for a few iterations at each step to
ensure that the actual minimum $\chi^2$ is found. For the latter,
we altered the normalization of the CXB spectra by 10\% (Kushino et
al. 2002) to approximate its field-to-field variation, and
similarly, re-normalized the NXB components by 2\%, 2\%, and 1\% for the ACIS, EPIC,
and XIS data (e.g., Hickox \& Markevitch 2006; De Luca \& Molendi 2004; Tawa et al. 2008),
respectively, to assess the error ranges. We
have also tried assigning a systematic error of 2\% to approximate
the unassessed calibration uncertainties of the EPIC data (e.g., Takahashi et al. 2009), which turned out to have negligible
impacts on the results and thus was not included in the actual fittings.

\section{ANALYSIS AND RESULTS}

\subsection{Azimuthally Averaged Spectral Analysis}
\subsubsection{Single-phase ICM Model}

We extracted the ACIS and EPIC spectra from eight thin annuli,
  i.e.,
$0-30$$h_{71}^{-1}$ kpc ($0^{\prime}-0^{\prime}.4$),
$30-51$$h_{71}^{-1}$ kpc ($0^{\prime}.4-0^{\prime}.7$),
$51-80$$h_{71}^{-1}$ kpc ($0^{\prime}.7-1^{\prime}.1$),
$80-116$$h_{71}^{-1}$ kpc ($1^{\prime}.1-1^{\prime}.6$),
$116-238$$h_{71}^{-1}$ kpc ($1^{\prime}.6-3^{\prime}.3$),
$238-354$$h_{71}^{-1}$ kpc ($3^{\prime}.3-4^{\prime}.9$),
$354-707$$h_{71}^{-1}$ kpc ($4^{\prime}.9-9^{\prime}.8$) and
$707-1335$$h_{71}^{-1}$ kpc ($9^{\prime}.8-18^{\prime}.5$), and the XIS
spectra from four thick annuli, i.e., $0-144$$h_{71}^{-1}$ kpc ($0^{\prime}-2^{\prime}.0$),
$144-320$$h_{71}^{-1}$ kpc ($2^{\prime}.0-4^{\prime}.4$),
$320-700$$h_{71}^{-1}$ kpc ($4^{\prime}.4-9^{\prime}.7$) and
$700-1444$$h_{71}^{-1}$ kpc ($9^{\prime}.7-20^{\prime}.0$). When
fitting these spectra, the lower
energy cut was set at 0.7 keV, 0.7 keV, and
0.6 keV for the ACIS, EPIC, and XIS, respectively, while the
upper cut was fixed at 8.0 keV for all the three 
  detectors. Also, the Si K edge
($1.8-1.9$ keV) was excluded from all the XIS spectra. Independent
fittings were carried out for the {\em Chandra} (ACIS-S and ACIS-I), {\em XMM-Newton}
(MOS and pn), and \suzaku\ (BI and FI) spectrum sets, by applying
a common absorbed APEC model and linking all annuli with {\tt XSPEC}
model PROJCT, which performs projection of 3-D shells onto 2-D
  annuli to evaluate the projected emission of outer shells on inner
  ones. For each shell, the gas temperature, metal abundance, and column density of the
neutral absorber were set free. The best-fit model yielded $\chi^{2}/\nu = 1750/1496$, 1040/881,
and 845/775 for the ACIS, EPIC, and XIS spectra, respectively.
Figure 2{\it b} and 2{\it c} show the best-fit deprojected 1P fittings 
to the spectra extracted from $320-700$$h_{71}^{-1}$ kpc and core
regions ($0-80$$h_{71}^{-1}$ kpc for the ACIS and EPIC, 
and $0-144$$h_{71}^{-1}$ kpc for the XIS), respectively.

As shown in Figure 3{\it a}, the best-fit deprojected 1P temperature drops inwards 
from $\approx 6.0$ keV at $\approx 100-350$$h_{71}^{-1}$ kpc to $\approx
3.0$ keV in the central 30$h_{71}^{-1}$ kpc, which shows apparent
diagnostic of a cool core (e.g., Sanderson et al. 2006). On the other
hand, in the cluster's outskirt ($\approx 350-1100$$h_{71}^{-1}$
kpc), the temperature declines outwards steeply down to $\approx
3.0$ keV. Generally, our 1P temperature profiles are consistent with
previous reports (e.g., Ettori et al. 2002; Vikhlinin et al.
2006; Snowden et al. 2008; Bautz et al. 2009). Incidentally, in
$\approx 150-300$$h_{71}^{-1}$ kpc, the temperature obtained with the
ACIS spectra is by about $1.0$ keV higher than those measured with the EPIC and XIS
spectra. This discrepancy can be ascribed to a local
temperature structure (see \S3.2.2), which is smoothed to some extent in the \xmm\
and \suzaku\ data.

Although the best-fit 1P model can reproduce the three data groups
with reasonable goodness, i.e., $\chi^{2}/\nu \approx 1.09 \sim 1.18$, the 
fitting residuals become drastically significant in the central 80$h_{71}^{-1}$ kpc
regions, where the averaged gas temperature is measured to be $\simeq
4.5$ keV (Fig. 3{\it a}). As shown in Figure 2{\it c}, the 1P model significantly underestimates the
Fe-L blend and the continuum in $>$ 5.0 keV. This indicates that these
annular spectra exhibit stronger multi-temperature nature than is
predicted by foreground and background contributions from different
radii that are accounted for by the PROJCT model. 
To further examine the issues with the 1P ICM model for these regions,
following, e.g., Cavagnolo et al. (2008), we fitted the ACIS spectra in $0.7-4.0$ keV and $4.0-8.0$ keV with the
1P model, both extracted from the central 80$h_{71}^{-1}$ kpc
and corrected for projection effects. The
1P gas temperature obtained in $0.7-4.0$ keV ($T_{\rm 0.7-4.0 \ keV} = 3.5\pm0.2$ keV) became lower than its counterpart
in $4.0-8.0$ keV ($T_{\rm 4.0-8.0 \ keV} = 5.7\pm0.7$ keV) at the 99\% confidence
level. The bandpass dependence cannot be ascribed to the
calibration problem of {\em Chandra} at soft band, because a similar
dependence can be obtained using the {\em XMM-Newton} data, with
$T_{\rm 0.7-4.0 \ keV} = 3.3\pm0.1$ keV and $T_{\rm 4.0-8.0 \ keV} = 4.7\pm0.5$ keV. Thus, the
1P modeling of the ICM is considered inadequate, and a 
2P ICM spectral model is hence invoked.

\subsubsection{Two-phase ICM Model}

Next we
investigate whether or not the 2P ICM model, as inferred in the above 1P
analysis, is applicable to the spectra
extracted in the inner 80$h_{71}^{-1}$ kpc region, after the projection
effects are taken into account. Here, the 2P ICM components were both
represented by thermal APEC models, which were constrained to have the same
metallicity and absorption. Both the cool- and hot-phase ICM
temperatures ($T_{\rm c}$ and $T_{\rm h}$, respectively) were tied
among all the thin shells in central 80$h_{71}^{-1}$ kpc, while the
1P ICM model (\S 3.1.1) was 
retained for the outer parts. This fitting yielded overall
$\chi^{2}/\nu$ = 1608/1494 and 929/879 for the ACIS and
EPIC spectrum sets, respectively. The fitting goodness was compared to the 1P one (\S 3.1.1) 
using an F-test, which yielded $F$-statistic of 66.0 and
52.5 for the ACIS and EPIC data, respectively, indicating that the 2P
model gives a better fit than the 1P model at $>$ 99\% confidence
level. As shown in Figure 2{\it
  c}, the 2P ICM model apparently better reproduces the Fe-L 
blend and the continuum in $>$ 5.0 keV than its 1P
counterpart. This shows that the 2P ICM model is
strongly and consistently required for the inner 80$h_{71}^{-1}$ kpc
region even after removing (via PROJCT) foreground and background
contributions from the outer shells.   
The 2P temperatures
($T_{\rm c}$, $T_{\rm h}$) were determined
as ($2.2 \pm 0.2$, $5.7 \pm 0.4$) keV and ($2.0 \pm 0.2$, $5.0 \pm
0.3$) keV by using the ACIS and EPIC data, respectively. For the
regions with $r > 80$$h_{71}^{-1}$ kpc, the 2P model does not
significantly improve the fitting over the 1P model.

The same 2P ICM model was also applied to the thick shell spectra
obtained with the XIS. Compared to the 1P case (\S 3.1.1; $\chi^{2}/\nu
\approx 845/775$), the 2P ICM model again gave
significantly better fit to the XIS spectra within central
$320$$h_{71}^{-1}$ kpc region, with reduced chi-squared improved to
$\chi^{2}/\nu$ = 802/773. Since the PSF of the XIS is much wider than
those of the ACIS and EPIC, with the XIS alone we cannot constrain the
boundary between the 1P and 2P regions to a small radius, as we did
with the ACIS and EPIC data. To examine the PSF effect on the
spectral fitting of the innermost thick shell ($0-144$$h_{71}^{-1}$
kpc), we performed a ray tracing simulation (e.g., Ishisaki et al. 2007;
Reiprich et al. 2009) based on the ACIS image, and found that the
$0.5-8.0$ keV counts in this shell are mostly ($\approx 75$\%) from
the emission originated in $0-80$$h_{71}^{-1}$ kpc, only a small
fraction ($\approx 5$\%) come from $116-238$$h_{71}^{-1}$ kpc region
where a hotter component ($T_{\rm X}\approx 6.5$ keV; Figure
3{\it a}) is detected with the
ACIS data. Hence this hot component cannot bias the 2P result significantly. In fact, the best-fit 2P temperatures
for this region, $T_{\rm c} = 2.1
\pm 0.5$ keV and $T_{\rm h} = 5.5 \pm 0.4$ keV, are consistent with those obtained in
the thin shell analysis. Thus, the PSF of the XIS little affects the 2P
results obtained for the innermost thick shell.

As the most consistent form of our 2P analysis, we simultaneously
fitted the thin shell spectra (the ACIS and EPIC) and those from the
thick shell (the XIS), by allowing to vary independently the pair of 2P temperatures of
each inner shell, i.e., $\leq 
80$$h_{71}^{-1}$ kpc and $\leq 320$$h_{71}^{-1}$ kpc for the ACIS/EPIC and XIS
cases, respectively. Nearly the same fit
goodness was achieved (Table 2), and the best-fit $T_{\rm c}$ and $T_{\rm h}$, as
shown in Figure 4{\it a} and Table 2, exhibit nearly insignificant spatial variations across the inner
shells. The values of $T_{\rm c}$ and $T_{\rm h}$ are consistent
among the three instruments, and broadly consistent with the \asca\ result, ($T_{\rm c}$, $T_{\rm h}$) =
($1.7 \pm 0.3$, $6.5 \pm 0.6$) keV, reported in Xu et
al. (1998). In short, the ICM in the cool core of this cluster can be described by
two discrete temperatures, $T_{\rm c}\approx 2.0-2.2$ keV and $T_{\rm
  h}\approx 5.0-5.7$ keV, which are both consistent with being spatially
constant. This makes the simple 2P ICM picture (Makishima et
al. 2001) a natural and reasonable description.

\subsubsection{A Weak 0.8 keV Spectral Component Within the Central
  144$h_{71}^{-1}$ kpc}

Employing the 1P formalism, Fabian et al. (2001) reported a
filamentary structure in the inner 50$h_{71}^{-1}$ kpc region of
A1795, with a possible temperature of $\sim 1$ keV, apparently lower
than the value of $T_{\rm c}$ obtained in our 2P analysis. To look for such a
component, we added a third APEC component to the 2P
ICM model describing the XIS data from the central 144$h_{71}^{-1}$ kpc
region. Indeed, we obtained a significantly better fit ($\chi^{2}/\nu =
791/771$) by adding a third APEC component, whose temperature is
$kT_{\rm 3} =  0.8$$\pm 0.4$ keV, while its metal abundance, 
absorption, and redshift were tied to those of the 2P ICM components.    
An $F$-test indicates that the probability of this improvement being caused by chance is
$< 5 \times 10^{-3}$. The $0.3-10.0$ keV luminosity of this 0.8
keV component is $3.6^{+1.2}_{-2.6}\times10^{42}$ ergs
$\rm s^{-1}$, which is consistent with the
luminosity of the filamentary structure measured with the ACIS in
Fabian et al. (2001; $\sim 4\times10^{42}$ ergs $\rm s^{-1}$). 
As shown in Table 2, the values of $T_{\rm c}$ and $T_{\rm h}$, as
well as the ICM abundances, remain 
nearly the same by adding this 0.8 keV component, because it
contributes only $\approx 0.7$\% of the total $0.3-10.0$ keV
luminosity of the central 144$h_{71}^{-1}$ kpc region ($\approx 5.0\times10^{44}$ ergs $\rm s^{-1}$).

In order to crosscheck the XIS result, we
analyzed the \xmm\ RGS and deprojected EPIC spectra extracted from the central
80$h_{71}^{-1}$ kpc region by fitting them with three component
spectral model, i.e., 2P ICM plus a
third APEC component. Since the RGS is not so sensitive to hot
gas components, we fixed $T_{\rm c}$ and $T_{\rm h}$ at the best-fit
EPIC results, i.e., 2.0 keV and 5.0 keV, respectively. By fitting the
RGS1 and RGS2 spectra simultaneously in $6-23$~\AA , a
weak 0.8 keV component was again detected, with a probability of $6
\times 10^{-2}$ for the detection to be caused by chance. The fitting
of the EPIC spectra was improved to $\chi^{2}/\nu =
923/878$ by including the 0.8 keV component, significantly better than
previous 2P fitting in terms of $F$-test ($>98$\% confidence level).
The $0.3-10.0$ keV luminosities 
of this component determined with the RGS and the EPIC are $2.7 \pm
2.3\times10^{42}$ ergs $\rm s^{-1}$ and $2.6 \pm 1.7\times10^{42}$
ergs $\rm s^{-1}$, respectively, nicely consistent with the XIS value
and the ACIS result in Fabian et al. (2001).

\subsubsection{Filling Factor of the Cool Phase}
Using the high resolution {\it Chandra}, {\it XMM-Newton}, and {\it
  Suzaku} data, we have demonstrated that the 2P ICM model gives a
significantly better description of the ICM thermal condition in the central
80$h_{71}^{-1}$ kpc of A1795 than the 1P counterpart. The cool and hot
phase ICM components have $0.3-10.0$ keV luminosities of about $1.4\times10^{44}$ ergs $\rm
s^{-1}$ and $2.5\times10^{44}$ ergs $\rm s^{-1}$, respectively. Also a
weak 0.8 keV component has been detected in the same region, which
has a $0.3-10.0$ keV luminosity of about $3.6\times10^{42}$ ergs $\rm s^{-1}$.

The 2P ICM picture implicitly assumes that these
phases with different temperatures coexist, each occupying a certain
fraction of the total volume under study in the cluster center, which can be
described as volume filling factor (e.g., Ikebe et al. 1999; Makishima et al. 2001;
Takahashi et al. 2009). To quantify this quantity, we followed the
method described in Ikebe et al. (1999) and 
calculated the filling factor of the cool phase gas, $\eta_{\rm
    c}(R)$, where $R$ denotes 3-D radius. With $\eta_{\rm c}$, the specific emission measures of the two
phases, $Q_{\rm c}(R)$ and $Q_{\rm h}(R)$, are described as
\begin{equation} \label{eq:em}
Q_{\rm c}(R) = n_{\rm c}(R)^{2}\eta_{\rm c}(R), \mbox{ } Q_{\rm h}(R) = n_{\rm h}(R)^{2}(1-\eta_{\rm c}(R)),
\end{equation}
where $n_{\rm c}(R)$ and $n_{\rm h}(R)$ are the density distributions
of the cool and hot
phases, respectively. Assuming a pressure balance between the two phases
\begin{equation} \label{eq:pb}
n_{\rm c}(R)T_{\rm c}(R) = n_{\rm h}(R)T_{\rm h}(R),
\end{equation}
we have
\begin{equation} \label{eq:vf}
\eta_{\rm c}(R) = \left[ 1+\left(\frac{T_{\rm h}(R)}{T_{\rm c}(R)}\right)^{2}
\left(\frac{Q_{\rm h}(R)}{Q_{\rm c}(R)}\right) \right] ^{-1}.
\end{equation}
Figure 4{\it b} shows radial profiles of $\eta_{\rm c}(R)$, calculated
with the best-fit deprojected ACIS, EPIC, and
XIS model parameters as listed in Table 2. Thus, the ACIS and EPIC datasets consistently
indicate that the hot component dominates in volume over its cool
counterpart not only in outer regions, but also in the core region
($<80$$h_{71}^{-1}$ kpc). The cool phase gas occupies up to $\eta_{\rm c} \approx 0.2$ of the volume
in the central $30$$h_{71}^{-1}$ kpc, whereas $\eta_{\rm c}$
declines steeply to below 0.05 outside the central $80$$h_{71}^{-1}$
kpc region. The XIS result for the inner $144$$h_{71}^{-1}$ kpc
  is consistent with those of the ACIS and
EPIC, whereas the relatively large value of $\eta_{\rm c}$
indicated by the outer $144-320$$h_{71}^{-1}$ kpc XIS bin can be
attributed to the broad PSF of the XIS. In fact,
by performing a ray tracing
simulation (see \S3.1.2 for more details), the outer XIS
bin at
$144-320$$h_{71}^{-1}$ kpc was found to be contaminated by photons
scattered from the inner regions, in such a way that $\approx 37$\% of the emission
in this region actually comes from the central 144$h_{71}^{-1}$
kpc. In this case, a weighted mean of $\eta_{\rm c} \approx 0.08$
measured with the XIS over inner $144$$h_{71}^{-1}$ kpc and $\eta_{\rm
  c} \approx 0.01$ measured with the other two missions at the
$144-320$$h_{71}^{-1}$ kpc region becomes $\approx 0.04$, in agreement
with the outer XIS point. Hence, after correcting for the PSF effect, all
data indicate the same 2P configuration in the central region.

\subsubsection{2P vs. Multi-phase ICM Model}

The ICM in the central $144$$h_{71}^{-1}$ kpc may alternatively in a
multi-phase condition (e.g., Kaastra et al. 2004), to be described by
a more complicated emission measure distribution in a wide temperature
range than that used in \S 3.1.2. To examine this possibility, we
analyzed the deprojected XIS spectra from the inner 
144$h_{71}^{-1}$ kpc region with the same multi-temperature fitting
approach as used in Tamura et al. (2001) and Takahashi et al. (2009). That is, the spectra
were modeled as a cumulative contribution of seven APEC components, whose
temperatures are given as $T_{\rm 0}$, 1.5$T_{\rm 0}$,
(1.5$)^2$$T_{\rm 0}$,..., and (1.5$)^6$$T_{\rm 0}$, where $T_{\rm 0}$
is a base temperature and left free in the fitting. The seven components
were constrained to have the same metal
abundance, and suffer the same absorption. The fit has been
acceptable, and the goodness of fitting ($\chi^{2}/\nu=
786/769$) is as good as that of
the 2P ICM plus a 0.8 keV component model (Table 2). The base 
temperature was constrained as $T_{\rm 0} = 0.7 \pm 0.2$ keV.  As
shown in Figure 5, in the multi-phase model, only one
weak component at $\approx 0.7$ keV and
two strong components at 2.4 keV and 5.3 keV remain significant (68\%
confidence level), while the rest components cannot be
constrained. This is essentially identical to the 2P ICM plus 
0.8 keV modeling. Our result is consistent with the multi-phase fitting 
with the EPIC data in Kaastra et al. (2004), which also shows a two-temperature structure
(2.5 keV and 4.9 keV, see their Table 6) in the cluster center. Hence, the XIS data 
prefers the discrete 2P model,
to a continuous temperature distribution, although the latter cannot
be ruled out based on available data.

\subsubsection{Metal Abundance and Absorption Distributions}

As shown in Figure 3{\it b}, the deprojected metal abundance profiles appear roughly consistent
between the 1P and 2P ICM modelings (\S3.1.1 and \S3.1.2,
respectively). Both profiles show a peak
in the shell of $30-51$$h_{71}^{-1}$ kpc, and a mild decline outwards.
This abundance profile, however, should be regarded as an average among
  those of different elements, because we have so far assumed the
  solar abundance ratios. To examine the ICM for possible deviation
  from the solar ratios, we employed two VAPEC models and reran
the deprojected 2P fittings of the \suzaku\ XIS
spectra extracted from the central 320$h_{71}^{-1}$ kpc
region. Specifically, we left the abundances of O, Mg, Si and Fe free,  
fixed the abundances of He, C and N at the solar value (Anders
\& Grevesse 1989), and tied the abundance of Ni to that of Fe and those of other
elements (Ne, Al, S, Ar and Ca) to that of Si. This model gave nearly
the same set of temperatures for the 2P ICM, and a fit 
goodness ($\chi^{2}/\nu=787/761$) slightly better than that of the best-fit
APEC model ($\chi^{2}/\nu=802/773$). As shown in Table 3, the best-fit Fe and Si
abundances increase significantly towards the center, while the O and Mg abundances
are nearly constant throughout the cluster. This confirms the previous
ACIS result reported in Ettori et al. (2002; see also Matsushita et al. 2007). 
We also added another APEC model ($A$ = 0.5
$Z_\odot$; see Table 2) in the central 144$h_{71}^{-1}$
kpc to account for the 0.8 keV component, while the model gave
nearly the same best-fit abundance profiles for the 2P ICM components.

To examine the possible existence of any intrinsic
absorption, we compared the absorption obtained in the spectral analysis
with the Galactic value. Regardless of the 1P or 2P modeling, the ACIS
and XIS fitting results revealed a significant excess absorption by
$1-2 \times 10^{20}$ $\rm cm^{-2}$ (Table 3)
beyond the Galactic value of $1.2 \times 10^{20}$ $\rm cm^{-2}$, at least
in the central 30$h_{71}^{-1}$ kpc region. The excess absorption could not be mitigated by
varying the abundances of specific elements (e.g., oxygen) whose emission lines couple with the
absorption feature. This result is in good agreement with the ACIS result of
Ettori et al. (2002; $\approx 2.5 \times 10^{20}$ $\rm cm^{-2}$). On
the contrary, our best-fit EPIC result ($\approx 
1.1 \times 10^{20}$ $\rm cm^{-2}$), as shown in Table 3,
implies lack of excess absorption, in agreement with Nevalainen et
al. (2007) who used
the same \xmm\ MOS data. The discordance among the three detectors
maybe caused by a temporal solar wind charge exchange emission, which
biased the EPIC absorption low. As
shown in Wargelin et al. (2004), the brightness of charge exchange can
reach $\sim 2\times10^{-6}$ photon $\rm s^{-1}$ $\rm arcmin^{-2}$
$\rm cm^{-2}$ in $0.5-0.9$ keV, which is sufficient to undermine the
absorption by $0.5-1.0 \times 10^{20}$ $\rm cm^{-2}$. The origin of
the central excess absorption detected with the ACIS and XIS, on the
other hand, still remains unclear.

\subsection{Projected 2-D Spectral Analysis}

As shown above, the azimuthally-averaged spectral analysis
  prefers a view of the 2P ICM in the central
80$h_{71}^{-1}$ kpc of A1795 than the 1P counterpart. However,
we cannot exclude at present the possibility that the 2P preference is
artificially caused by an
anisotropic spatial distribution of 1P gas temperature that fluctuates
between the range from $T_{\rm c}$ to $T_{\rm h}$. To address this issue, we examine
two-dimensional (2-D) temperature distribution in the central region
of A1795. Below, 2-D ICM
abundance and $Q_{\rm c}$/$Q_{\rm h}$ 
distributions are also presented.

\subsubsection{Analysis Procedure}
Combining five \chandra\ ACIS-S pointings onto the central region
together, we have collected sufficient photons for a high resolution 2-D spectral
analysis. Following the procedure described in detail in Gu et
al. (2009), a set of $> 10000$ discrete points (${\bf r}_{\rm i}$, $i
= 1,2,3,...$), or ``knots'', were chosen  
in the central
$240$$h_{71}^{-1}$ kpc, which are randomly distributed with a
separation of $<6$$h_{71}^{-1}$ kpc between any two adjacent knots. To
each knot we assigned a circular region, which is centered on the knot
and has an adaptive radius of $15-80$$h_{71}^{-1}$ kpc, ensuring that it
encloses $>10000$ photons in $0.5-8.0$ keV after all detected point
sources (\S2.2)
were excluded. The spectrum extracted from each
circular region was fitted with the 1P and 2P APEC models, both
subjected to an absorption that was set free. In the
2P spectral fitting, the temperatures of the cool and hot
components were fixed at 2.2 keV and 5.7 keV (i.e., best-fit ACIS
results; \S 3.1.2), respectively,
and the two components were assumed to have the same abundance and
absorption. For each knot ${\bf r}_{\rm i}$, we obtained the best-fit 1P gas
temperature, 1P/2P abundance,
and specific emission measure ratio between the cool and hot
components ($Q_{\rm c}$/$Q_{\rm h}$), as well as their $1\sigma$
errors. 

Then, following Gu et al. (2009), we calculated continuous maps based
on the obtained knots. For any position ${\bf r}$ within the map
region, we defined a scale 
$s({\bf r})$, so that there are $>10000$ net photons in a circular
region centered at ${\bf r}$, whose radius is $s({\bf r})$. The 1P
temperature at ${\bf r}$ was then calculated by a weighted mean of all
the knots ${\bf r}_{\rm i}$ in the circular region, 
   \begin{equation} \label{eq:tmap}
T({\bf r}) = \sum_{\bf r_i} (G_{\bf r_i}(R_{{\bf r},{\bf r_i}})T_{\rm c}({\bf r_i}))/\sum_{\bf r_i} G_{\bf r_i}(R_{{\bf r},{\bf r_i}}), \mbox{ } \rm when \mbox{ } \it R_{{\bf r},{\bf r_i}} < s({\bf r}),
\end{equation}
where $R_{{\bf r},{\bf r_i}}$ is the distance from ${\bf r}$ to ${\bf r_i}$, and $G_{\bf r_i}$ is the Gaussian kernel whose scale
parameter $\sigma$ is fixed at $s({\bf r_i})$. The use of compact
Gaussian kernel guarantees an angular resolution of $\sim
15$$h_{71}^{-1}$ kpc within central 100$h_{71}^{-1}$ kpc region. The
abundance and $Q_{\rm c}$/$Q_{\rm h}$ maps, along with their $1\sigma$ error maps, are calculated in the same
way. The resulting 1P temperature and abundance maps are shown in
Figure 6, and the 2P abundance (hereafter $A_{\rm 2P}$) and $Q_{\rm
  c}$/$Q_{\rm h}$ maps are presented in Figure 7. The 0.8
keV component (\S 3.1.3) was not considered in the above 2-D analysis,
because it will introduce negligible effects, i.e., uncertainties of
about 0.05 keV and 0.02 $Z_\odot$, to the temperature and abundance
measurements, respectively.

\subsubsection{Relaxed Cool Core and SE High Temperature Arc}

The primary purpose of 2-D spectral analysis is to assess the validity
of spherically symmetric temperature distribution that we assumed in the radial 1P/2P analysis. To do
this, we examined the obtained 1P temperature map for any significant
anisotropic distribution. As shown in Figure 6{\it a}, the 1P temperature
distribution in the cluster central region exhibits an approximate elliptical
symmetry; the temperature variation in the azimuthal direction is
about 0.5 keV, 1.0 keV, and 1.0 keV at $r=20$$h_{71}^{-1}$ kpc,
$40$$h_{71}^{-1}$ kpc, and $80$$h_{71}^{-1}$ kpc, respectively, which
is insufficient, at least by a factor of three, to imitate the obtained 2P ICM condition. Similar morphology
is seen in the $Q_{\rm c}$/$Q_{\rm h}$ map of the core region as shown
in Figure 7{\it b}. This confirms that the 2P view is not an artifact caused
by a 1P ICM with large azimuthal asymmetry. It also indicates that the
two phases, in terms of the 2P view, must be separated on scales smaller than the
spatial resolution allowed by the present analysis ($\approx 15$$h_{71}^{-1}$ kpc).

The most prominent feature on the 1P temperature map (Fig. 6{\it a}) is a high-temperature arc
($\approx 7.5$ keV according to the 1P model) located at $\approx 100 - 180$$h_{71}^{-1}$ kpc
southeast (SE) of the cluster's center, with an open angle of $>$$120
^{\circ}$. Given the temperature difference of about 2 keV, the hot structure is significant
over the ambient on 90\% confidence level. This feature agrees
with a high temperature bin ($116-238$$h_{71}^{-1}$ kpc) on the deprojected ACIS temperature
profile shown in Figure 3{\it a}. A same high temperature structure
was found by Markevitch et al. (2001), who carried out 
a 1P azimuthal spectral analysis for the southern part of the
cluster. Markevitch et al. (2001) also reported a density jump by a
factor of $1.3-1.5$ near the inner edge ($\approx 85$$h_{71}^{-1}$
kpc from the cluster center) of the high-temperature arc, and ascribed
the density jump to a cold front. Since both the high-temperature component 
and the cold front locate outsides of the 2P region (i.e., $r < 80$$h_{71}^{-1}$ kpc),  
they are unable to affect the obtained 2P result significantly, even for the 
XIS thick shell as indicated by the ray tracing simulation shown in \S 3.1.2.

\subsubsection{A Possible Correlation Between Metal-rich and Cool-phase Gas}

A comparison of the $A_{\rm 2P}$ map (Fig. 7{\it a}) with the
$Q_{\rm c}$/$Q_{\rm h}$ map (Fig. 7{\it b}) suggests
a spatial correlation on a scale of $50-100$$h_{71}^{-1}$ kpc
  between the two 
quantities. Both maps reveal strongly inhomogeneous
distributions outside the relaxed core region, with substructures protruding
towards southwest, northeast, and 
northwest of the cluster center by $\approx 80-100$$h_{71}^{-1}$ kpc.
Most of these substructures show both higher metal abundances,
typically by a
factor of $2-3$, and larger cool phase fractions,
than their neighborhoods. The best-fit $A_{\rm 2P}$ and $Q_{\rm
  c}$/$Q_{\rm h}$ values for all the knot-centered circular regions in
$50-100$$h_{71}^{-1}$ kpc, obtained in our 2-D spectral analysis (\S
3.2.1), were compared directly in Figure 7{\it c}. Indeed, it reveals a positive
linear correlation, which is represented by
an analytic form as $A_{\rm 2P} = 0.38^{+0.11}_{-0.12} \times Q_{\rm
  c}$/$Q_{\rm h}+ 0.48^{+0.08}_{-0.07}$ $Z_\odot$. The linear correlation
coefficient was obtained as 0.97.

To quantify the suggested correlation, following, e.g., Shibata et
al. (2001), we calculated a 2-D cross
correlation function as, 
 \begin{equation}\label{eq:tpcf}
\xi ({\rm R}) =
\left < \frac{\left \{ A_{\rm 2P}({\bf r}_{1}) - \overline{A_{\rm 2P}} \right \}
\left \{ Q_{\rm c}({\bf r}_{2})/Q_{\rm h}({\bf r}_{2}) - \overline{Q_{\rm c}/Q_{\rm h}}
\right \} }{\overline{A_{\rm 2P}}\overline{Q_{\rm c}/Q_{\rm h}}}
\right >_{\rm r_{12} = \rm R},
\end{equation}
where $\overline{A_{\rm 2P}}$ and $\overline{Q_{\rm c}/Q_{\rm h}}$ are averages
  of $A_{\rm 2P}$ and $Q_{\rm c}/Q_{\rm h}$ maps, respectively, $\rm
  r_{12}= |{\bf r}_{1} - {\bf r}_{2}|$ is the distance between ${\bf
      r}_{1}$ and ${\bf r}_{2}$, and the bracket is ensemble
    average. The obtained correlation function is shown in Figure 7{\it d}. 
A reference profile ($\xi_{\rm ref} ({\rm R})$) was also calculated, based on
    a series of random maps, $A_{\rm 2P, ran}$ and $Q_{\rm c,
      ran}/Q_{\rm h, ran}$, which were obtained by randomizing the maps within
    the same data range and smoothing to the same spatial resolution as the original $A_{\rm 2P}$ and $Q_{\rm
        c}/Q_{\rm h}$ maps. The error bars shown in Figure 7{\it d} were calculated from the
      variances of $\xi ({\rm R})$ when scattering the $A_{\rm 2P}$
      and $Q_{\rm c}/Q_{\rm h}$ maps by their $1\sigma$ error
      maps. Comparing to the $\xi_{\rm ref} ({\rm R})$, the $\xi ({\rm
        R})$ profile shows a significant excess within 100$h_{71}^{-1}$ kpc,
      reconfirming the result that cool-phase ICM is indeed more metal-rich.

Given the detected correlation, the assumption made in \S 3.1.2 and
\S3.1.6 that the two phases ICM have approximately the same
abundances may not stand now. Therefore, we refitted the
deprojected XIS spectra of the central $144$$h_{71}^{-1}$ kpc region
with the 2P ICM (VAPEC + VAPEC) + 0.8 keV component (APEC; 0.5
$Z_\odot$) model, same
as the one in \S3.1.6, except that the iron abundances of the
cool and hot phases were let float separately. The
absorption column densities of the three components were again tied together
in the fitting as a free parameter.  We found that the abundance of
cool phase ICM ($A_{\rm Fe, c}$) does appear higher 
than that of the hot one ($A_{\rm Fe, h}$); i.e., ($A_{\rm Fe, c}, A_{\rm Fe, h}$) =
($0.80 \pm 0.25$, $0.36 \pm 0.06$) $Z_\odot$. Compared to the
  model presented in \S3.1.6 (Table 3), the current model improves fit goodness
  to $\chi^{2}/\nu = 779/759$ from $787/761$,  with a probability of $2
\times 10^{-2}$ for the improvement to be caused by chance. 

Following,
e.g., Simionescu et al. (2008), the iron abundance to be obtained with
a 2P model assuming a single common metallicity can be estimated by 
\begin{equation} \label{eq:aew}
A_{\rm Fe}^{\prime} \approx \frac{Q_{\rm c}A_{\rm Fe, c}+Q_{\rm h}A_{\rm Fe, h}}{Q_{\rm c}+Q_{\rm h}}.
\end{equation}  
Eliminating $Q_{\rm c}$ and $Q_{\rm h}$ with Eq.(1),
we have
\begin{equation} \label{eq:avf}
A_{\rm Fe}^{\prime} \approx \frac{T_{\rm h}^2 \eta_{\rm c} A_{\rm Fe,
    c} + T_{\rm c}^2 (1-\eta_{\rm c}) A_{\rm Fe, h}}{T_{\rm h}^2
  \eta_{\rm c} + T_{\rm c}^2 (1-\eta_{\rm c})}. 
\end{equation}   
Adopting $\eta_{\rm c} = 0.08$ (Fig. 4{\it b}) and ($A_{\rm Fe, c}, A_{\rm Fe, h}$)
obtained above, $A_{\rm Fe}^{\prime}$ is then calculated as 0.51$Z_\odot$, which
agrees with that derived with the XIS spectra ($0.48\pm0.03$ $Z_\odot$, Table 3). Thus, all the ACIS
and XIS results obtained so far can be interpreted as evidence for the
relatively high metal abundance of the cool phase ICM.

\subsection{Hierarchical Gravitational Potential Structure}

\subsubsection{Central Excess in X-ray Surface Brightness}


In many cool-core clusters, the X-ray surface brightness exhibits an
central excess over a $\beta$ model that well fits the outer
region. Makishima et al. (2001) argued that the central excess is a combined
result of two major effects; the presence of a cool phase component, 
and the existence of hierarchical potential structure. According to their
definition, the hierarchical potential is specified as a halo-in-halo structure, i.e.,
a smaller potential component is nested on a larger one, so that the 
overall potential shows a central deepening relative to a simple King profile.
Here we examined the surface brightness profile of A1795 for such a central excess.
In order to measure the surface brightness profile precisely by taking 
advantages of both high spatial resolution of the \chandra\ ACIS and 
stable low background of the \suzaku\ XIS, we calculated
exposure-corrected surface brightness profiles from the inner 1000$h_{71}^{-1}$ kpc ACIS and
$1200-1800$$h_{71}^{-1}$ kpc XIS data, i.e., $S_{\rm ACIS}(r)$ and $S_{\rm
  XIS}(r)$, respectively, where $r$ is projected radius. The method
  described in, e.g., Markevitch et al. (1998), was employed to compensate
discrepancies on calibration and instrumental background between
the two instruments. First, we performed simulations using the {\tt xissim} tool (Ishisaki et
al. 2007) to smooth the {\em Chandra} images with the {\em Suzaku} PSF,
and extracted the simulated surface brightness profile in the central
1000$h_{71}^{-1}$ kpc, i.e., $S_{\rm ACIS}^{\prime}(r)$. Then, $S_{\rm
  ACIS}^{\prime}(r)$ was scaled to $S_{\rm
  XIS}(r)$ by solving $aS_{\rm ACIS}^{\prime}(r)-b=S_{\rm XIS}(r)$, where $a$ and
$b$ represent the differences in normalization and NXB, respectively. A modified 
$S_{\rm ACIS}(r)$ with the original ACIS resolution was obtained by applying $a$ and $b$ to $S_{\rm ACIS}(r)$
in the same way as we normalized $S_{\rm ACIS}^{\prime}(r)$. 

Then, the modified $S_{\rm ACIS}(r)$ and $S_{\rm XIS}(r)$, shown in Figure 8, were
fitted jointly with the $\beta$ model, $S(r)= S(0) \{1+(r/r_{\rm
  c})^2\}^{-3\beta+1/2} + S_{\rm BG}$, where $S(0)$ is the central
brightness, $r_{\rm c}$ is the core radius, and $S_{\rm BG}$ is the background.
To represent the possible halo-in-halo structure, 
a double-$\beta$ model was also tested as $S(r) = S_1(0) \{1+(r/r_{\rm c1})^2\}^{-3\beta_1 + 1/2} +
S_2(0) \{1+(r/r_{\rm c2})^2\}^{-3\beta_2 + 1/2}+ S_{\rm BG}$, where
subscripts 1 and 2 denote compact and extended components,
respectively. As shown in Table 4, the fit with the $\beta$ model can
be rejected on the 95\% confidence level, because it significantly
underestimates the surface brightness in central $\sim
100$$h_{71}^{-1}$ kpc, while the double-$\beta$ model gives an
acceptable fit to the data. The successful fits in Figure 8 are based
on a still more sophisticated model, a $\beta$+double-$\beta$ gas density model,
to be explained later. The properties of the central excess
  brightness, represented by 
the compact $\beta$ component ($\beta_1 \approx 0.75$, $r_{\rm c1} \approx 53$$h_{71}^{-1}$ kpc), 
are consistent within errors between the
$0.5-3.0$ keV and $3.0-8.0$ keV bands. Such an energy-independent central 
excess is likely to be caused by an intrinsic hierarchical potential shape,
rather than the presence of cool phase component. This result agrees with
the earlier \asca\ result reported in Xu et al. (1998).

To further quantify the central emission excess, we corrected the surface brightness profile 
for the projection effect using the best-fit deprojected 2P gas temperature and abundance profiles (Table 2; \S 3.1.2). The 
observed surface brightness profile was modeled as  
\begin{equation}
S(r)= \int^{\infty}_{\rm r}\Lambda(T_{\rm c},A_{\rm c})Q_{\rm c}(R)\frac{RdR}{\sqrt{R^{2}-r^{2}}}
+ \int^{\infty}_{\rm r}\Lambda(T_{\rm h},A_{\rm h})Q_{\rm h}(R)\frac{RdR}{\sqrt{R^{2}-r^{2}}} + S_{\rm BG},
\end{equation}
where $\Lambda$ is the cooling function, $Q_{\rm c}$ and $Q_{\rm h}$
are the same as in Eq.(1), and $S_{\rm BG}$ is the averaged background value.
Following Ikebe et al. (1999), we represented the specific emission measure
profiles by a 2P $\beta$+double-$\beta$ model, which consists of a
$\beta$ component for the cool phase ICM,
\begin{equation}
Q_{\rm c}(R) = \left \{
\begin {array} {c}
Q_{\rm 0,c} \left[ 1 + (\frac{R}{R_{\rm c,c}})^{2} \right] ^{-3\beta_{\rm c}}
\mbox{ } \mbox{ } \mbox{ } \mbox{ } R \leq 80h_{71}^{-1}\mbox{ } {\rm kpc}, \\
0 \mbox{ } \mbox{ } \mbox{ } \mbox{ } \mbox{ } \mbox{ } \mbox{ } \mbox{ } \mbox{ }
\mbox{ } \mbox{ } \mbox{ } \mbox{ }  \mbox{ } \mbox{ } \mbox{ } \mbox{
} \mbox{ } \mbox{ } \mbox{ } \mbox{ } \mbox{ } \mbox{ } \mbox{ }
\mbox{ } \mbox{ } \mbox{ } \mbox{ } \mbox{ } \mbox{ }
 R > 80h_{71}^{-1}\mbox{ } {\rm kpc},
\end{array}
\right .
\end{equation}
and a double-$\beta$ component for the hot phase ICM,
\begin{equation}
Q_{\rm h}(R) =
           Q_{\rm 0,h1} \left[ 1 + \left(\frac{R}{R_{\rm c,h1}}\right)^{2} \right] ^{-3\beta_{\rm h1}}
         + Q_{\rm 0,h2} \left[ 1 + \left(\frac{R}{R_{\rm c,h2}}\right)^{2} \right] ^{-3\beta_{\rm h2}},
\end{equation} 
where $Q_{\rm 0,c}$, $Q_{\rm 0,h1}$, and $Q_{\rm 0,h2}$ are the model
normalizations of the cool phase, the compact hot phase, and the
extended hot phase components, respectively. The filling factor
$\eta_{\rm c}$ is renormalized into these parameters.

With these preparations, we fitted the $0.5-8.0$ keV ACIS + XIS surface
brightness profiles using Eq.(8), where $\Lambda$ was calculated from
the best-fit 2P spectral parameters. The fit has
been successful with $\chi^{2}/\nu = 161/151$. The best-fit values of
$R_{\rm c,c}$, $R_{\rm c,h1}$, $R_{\rm c,h2}$, $\beta_{\rm c}$,
$\beta_{\rm h1}$, and $\beta_{\rm h2}$ are given in Table 4. The fittings with surface
brightness profiles extracted in different energies (i.e., $0.5-3.0$ keV and
$3.0-8.0$ keV bands) are shown in Figure 8. As presented in Table 4, all the fittings 
yielded consistent parameters for the 2P $\beta$+double-$\beta$ model.

In some clusters, the central excess can be alternatively
explained by a cuspy dark matter distribution proposed in Navarro, Frenk, \& White (1996; 
hereafter NFW model). The NFW model predicts 
potential structure with a single spatial scale, instead of the dual structure assumed by
the $\beta$+double-$\beta$ model. Given the NFW dark matter density profile,
\begin{equation}
\rho_{\rm DM}(R) = \rho_{\rm c} \delta_{\rm c} \left( \frac{R}{R_{\rm s}}\right)^{-1} \left( 1 + \frac{R}{R_{\rm s}}\right)^{-2}, 
\end{equation}    
where $\rho_{\rm c}$ is the critical density, $\delta_{\rm c}$ is characteristic density, and $R_{\rm s}$
is scale radius, the ICM density profile is expressed as
\begin{equation}
n_{\rm NFW} (R) = n_{\rm 0,NFW} \frac{\left(1+ \frac{R}{R_{\rm s}}\right)^{\frac{\alpha R_{\rm s}}{R}}-1}{e^{\frac{R}{R_{\rm s}}}-1}, 
\end{equation}    
where $n_{\rm 0,NFW}$ is model normalization and $\alpha$ is a parameter related to the ICM temperature. Then
the projected ICM surface brightness profile was modeled as a single component, 
\begin{equation}
S(r)= \int^{\infty}_{\rm r}\Lambda(T_{\rm 1P},A_{\rm 1P})n_{\rm NFW}^{2}(R)\frac{RdR}{\sqrt{R^{2}-r^{2}}} + S_{\rm BG}.
\end{equation}
We found this NFW model cannot give acceptable fits to the $0.5-8.0$ keV surface brightness profile for 
the entire cluster, with a minimum $\chi^{2}/\nu = 307/155$. As shown in Figure 8, the central excess emission is 
too strong to be explained by the NFW model that best-fits the $>100$$h_{71}^{-1}$ kpc region. Hence, we reconfirmed the 
{\it ASCA} result reported in Xu et al. (1998) that the 
halo-in-halo hierarchical model is preferred to the single-halo NFW model in A1795.

\subsubsection{Central Excess in Total Gravitating Mass}

Next we calculated the total gravitating mass profile $M(R)$ based on the best-fit 
2P temperatures (\S3.1.2) and the emission measure profiles obtained with the $\beta$+double-$\beta$ model 
(\S3.3.1). With the help of Eqs.(1)-(3), the best-fit emission measure profiles, $Q_{\rm
  c}(R)$ and $Q_{\rm h}(R)$, were converted to gas density profiles of the cool
and hot phases, $n_{\rm c}(R)$ and $n_{\rm h}(R)$,
respectively. Then, under the assumptions of spherical symmetry
and hydrostatic equilibrium, the
gravitating mass profile $M(R)$ was calculated as
\begin{equation} \label{eq:totalmass}
M (R) = - \frac{R^2}{G \rho_{\rm g}(R)}
\frac{{\rm d}P (R)}{{\rm d}R},
\end{equation}
where $P(R)$ = $n_{\rm c}(R)k_{\rm B}T_{\rm c}(R)$ = $n_{\rm h}(R)k_{\rm B}T_{\rm h}(R)$ is
the gas pressure, $\rho_{\rm g}(R) = \mu m_{\rm p} \{ \eta_{\rm c}(R)
n_{\rm c}(R) + [1 - \eta_{\rm c}(R)] n_{\rm h}(R) \}$
is the averaged gas mass density, $\mu = 0.609$ is the approximated mean molecular weight, and $m_{\rm p}$ is the proton mass. As shown
in Figure 9, the obtained mass profile exhibits a significant shoulder-like feature at
$R_{\rm x} \approx 120$$h_{71}^{-1}$ kpc, enclosing an excess mass
($M_{\rm excess}$) of about $1.5 \times 10^{13} M_\odot$ above
a flat core mass distribution calculated with the extended hot phase
component alone. Our result agrees with the \asca\ result, $R_{\rm x}
\approx 110$$h_{71}^{-1}$ kpc and $M_{\rm excess} \approx 2.1\times
10^{13} M_\odot$, as reported in Xu et al. (1998).

To examine systematic errors of the mass profile due to
different modelings of the ICM temperature distribution, we also calculated
the gravitating mass profile using the best-fit deprojected 1P spectral
parameters (\S 3.1.1). The projected X-ray surface brightness profile was
  modeled with a single ICM component as
\begin{equation}
S(r)= \int^{\infty}_{\rm r}\Lambda(T_{\rm 1P},A_{\rm 1P})Q_{\rm
  1P}(R)\frac{RdR}{\sqrt{R^{2}-r^{2}}} + S_{\rm BG},
\end{equation}
where the specific emission measure is given by a double-$\beta$ model, 
\begin{equation}
Q_{\rm 1P}(R) = n^{2}_{\rm 1P}(R) = 
           Q_{\rm 0,1} \left[ 1 + \left(\frac{R}{R_{\rm c,1}}\right)^{2} \right] ^{-3\beta_{\rm 1}}
         + Q_{\rm 0,2} \left[ 1 + \left(\frac{R}{R_{\rm c,2}}\right)^{2} \right] ^{-3\beta_{\rm 2}}.
\end{equation} 
The fit is as good as the 2P case, with $\chi^{2}/\nu = 165/154$. The
resulting mass profile, derived by substituting $P(R)$ = $n_{\rm 1P}(R)k_{\rm B}T_{\rm 1P}(R)$ and
$\rho_{\rm g}(R) = \mu m_{\rm p} n_{\rm 1P}(R)$ to Eq.(14), is shown in
Figure 9 by a red curve. Thus, the shoulder-like potential
structure is again found, although the 1P model systematically underestimates the
gravitating mass within 80$h_{71}^{-1}$ kpc by $\approx 30$\%. This
bias is negligible in $>80$ kpc.

Another concern is that, the assumption of hydrostatic equilibrium may
not hold for the ICM near the cold front located southeast of
the cluster center (\S 3.2.2), casting doubt on the validity of mass profile
obtained in the related region (Markevitch et al. 2001). To
investigate this, we excluded the southern half of the cluster and
recalculated mass profile for the northern half using the best-fit 2P
spectral results and surface brightness profile with the ACIS data. The 2P
temperatures, $T_{\rm c, north} = 2.0$ keV and $T_{\rm h, north} =
5.9$ keV, are consistent with previous ACIS results. The same is the
obtained $n_{\rm h, north}(R)$ profile, except for the annulus of the cold front (i.e., $R = 100-180$$h_{71}^{-1}$ kpc), where $n_{\rm h, 
  north}(R)$ is lower than $n_{\rm h}(R)$ by 
$5-20$\% in $100-180$$h_{71}^{-1}$ kpc. After comparing $M_{\rm
  north}(R)$ to $M(R)$ in Figure 9,     
we found that the shoulder-like structure, though
diminished by $\approx 10$\%, still remains significant after 
the high-temperature arc is excluded. Hence we conclude that the
existence of central
hierarchical structure is unambiguously confirmed in A1795.

\subsection{Hard X-ray Emission Component}

Here we examined the $12.0-50.0$ keV HXD-PIN data for possible
existence of any extra emission component. First, following, e.g.,
Nakazawa et al. (2009), we calculated the effective area of the
HXD-PIN considering the location and extension of the source. Based on the observed
best-fit gas temperature and gas density profiles
(Fig. 4{\it a} and Table 4), we calculated the ICM surface brightness distribution
map $S_{\rm hard}(x,y)$ in $12.0-50.0$ keV. Then we divided the
whole HXD field of view into
$1^{\prime}\times1^{\prime}$ blocks, calculated the point source
ARF for each block at $(x,y)$ using the ftool hxdarfgen, and convolved
all the monochromatic ARFs with the normalized $S_{\rm hard}(x,y)$. The PIN/XIS
cross normalization factor of 1.132 (\suzaku\ Memo
2007-11\footnote{http://www.astro.isas.ac.jp/suzaku/doc/suzakumemo/suzakumemo-2007-11.pdf})
was also adopted in the resulting ARF. 
To assess the NXB level, we utilized the ``tuned'' NXB model, which
provides an optimized reproductivity by making use of HXD-GSO
information (\suzaku\ Memo
2008-03\footnote{http://www.astro.isas.ac.jp/suzaku/doc/suzakumemo/suzakumemo-2008-03.pdf}). The
CXB model was assumed as $N(E) = 8.69 \times 10^{-4} \times
(E/$1.0 keV$)^{-1.29} \times$ $\rm exp$$(-E/$40.0 keV) photons $\rm cm^{-2}$ $\rm s^{-1}$
$\rm keV^{-1}$, where the normalization was set to match the HXD-PIN opening
angle of 0.32 $\rm deg^{2}$ (e.g., Nakazawa et al. 2009).

As shown in Figure 10, the NXB-subtracted XIS and HXD-PIN spectra in
$4-50$ keV were
tentatively fitted with a CXB + ICM model. The CXB component was
modeled by a cutoff powerlaw with index of 1.29 (Boldt 1987), and
the ICM component was represented by the 2P model derived in \S3.1.2.   
The 2P temperatures were set to ($T_{\rm c}$, $T_{\rm h}$) = ($2.1$, $5.5$) keV, and
their relative normalizations were fixed at the best-fit value in our
2P analysis with the XIS. The fitting is acceptable with $\chi^{2}/\nu
= 45/55$, and no additional hard X-ray component is required in
$12-50$ keV band. To obtain an upper limit on any
hard X-ray component, we fitted the combined XIS-HXD spectra with a 2P
ICM plus $\Gamma=2$ power-law model (Nakazawa et al. 2009). Taking
account of the NXB 
systematic uncertainty of 2.0\% at 90\% confidence level (Fukazawa et
al. 2009), the upper
limit flux of power-law emission in
$12.0-50.0$ keV band was estimated to be $8.2 \times 10^{-12}$ ergs
$\rm cm^{-2}$ $\rm s^{-1}$. A similar upper limit on any thermal hard X-ray
excess, $6.2 \times 10^{-12}$ ergs $\rm cm^{-2}$ $\rm s^{-1}$, was obtained by  
alternatively fitting the combined spectra with 2P ICM plus a 10 keV APEC
model.

\section{DISCUSSION}

We have analyzed {\em Chandra}, {\em XMM-Newton}, and {\em Suzaku}
data of the relaxed galaxy cluster A1795. The deprojected spectra extracted
from the central 80$h_{71}^{-1}$ kpc region have been successfully
reproduced by invoking two major ICM components with discrete
temperatures, indicating a 2P property. A third weak 0.8 keV component
is marginally detected in the core region. In \S3.2, we analyzed the
{\em Chandra} data, under 2-D (spatially) and 2P (spectroscopically)
formalism. The ICM metallicity was found to show a similar spatial
distribution to that of the cool phase 
component, indicating that the cool phase ICM is more metal-enriched than
the hot phase one.


\subsection{Two-phase ICM Properties}

A joint fitting of the thin shell spectra from the ACIS and EPIC,
and thick shell spectra from the XIS, consistently indicates a clear
preference of the
two-phase ICM nature, over the single-phase one, in the central
80$h_{71}^{-1}$ kpc region of A1795. The ICM therein can be
characterized by two representative temperatures, $T_{\rm h} = 5.0 -
5.7$ keV and $T_{\rm c} = 2.0 - 
2.2$ keV for the hot and cool phases, respectively, whose $0.3-10.0$
keV luminosities were estimated to be $L_{\rm h} = 2.5\times10^{44}$
ergs $\rm s^{-1}$ and $L_{\rm c} = 1.4\times10^{44}$ ergs $\rm
s^{-1}$. 
Both components show
insignificant temperature gradients in the 80$h_{71}^{-1}$ kpc region. The hot phase ICM
dominates in volume over the cool phase one by a factor of $\geq 4$,
even in the innermost region. The 2P view is consistent with
previous \asca\ results reported in Xu et al. (1998). Employing the
2P ICM model and fitting the cool phase and hot phase components with
$\beta$ and double-$\beta$ gas density models, respectively, we
quantified the shape of cluster potential. As shown in Figure 9, the
total gravitating mass profile exhibits a
central mass excess ($M_{\rm excess} \approx 1.5 \times 10^{13}
M_\odot$), which has a similar spatial scale to the region of
clear 2P property. This indicates a
potential hierarchy, with a 100$h_{71}^{-1}$ kpc level halo nested in
the center of a cluster-scale one.



 
As shown in Figure 3{\it b}, the ICM metallicity is also enhanced in the 2P region.
Furthermore, by comparing the 2-D metallicity and the emission measure ratio maps
created with the 2P ICM model, we found a possible spatial
correlation on $\geq 1\sigma$ level between the cool-phase fraction
(or $\eta_{\rm c}$) and the ICM metallicity in the $50-100$$h_{71}^{-1}$ kpc region. The 2-D maps
also revealed a filamentary distribution of the cool, metal-rich gas in
this region. The subsequent
2P fitting to the XIS spectra confirmed the metallicity enhancement
in the cool phase relative to the hot phase, giving the best-fit
iron abundances of the cool phase and hot phase ICM to be 0.80 $Z_\odot$
and 0.36 $Z_\odot$, respectively. Such a picture resembles the
previous findings of the metal-rich, multi-phase arms in M87
(Simionescu et al. 2008). In the innermost core of A1795, i.e., $r <
25$$h_{71}^{-1}$ kpc, however, a dip is seen in the abundance
distribution, which does not correlate with the cool phase
fraction. Such 
a feature is also seen in the coolest spots of some other clusters,
while its origin still remains unclear (e.g., Sanders et al. 2006).

In \S 3.1.3 and \S 3.1.5, we showed that the XIS and RGS spectra of the
central 144$h_{71}^{-1}$ 
kpc region can be better reproduced by including an additional weak 0.8 keV
component, without significantly revising the properties of the cool and hot ICM
components. This 0.8 keV component has a
$0.3-10.0$ keV luminosity of $3.6^{+1.2}_{-2.6}\times10^{42}$ ergs
$\rm s^{-1}$, and is dimmer than the hot ICM component by two orders
of magnitude. Since the temperature and luminosity of this component
are consistent with those of an X-ray/H$\alpha$ filament in the core
of A1795 as reported in Fabian et al. (2001), the 0.8 keV emission might be
associated with a portion of ISM of the cD galaxy, which is presumably  
confined within some filamentary magnetic structures.   


The 2P formalism has also been successful for the central region of
the Centaurus cluster, as firstly reported by Fukazawa et al. (1994)
and Ikebe et al. (1999) with {\it ASCA} data, and recently reinforced
by Takahashi et al. (2009) using {\it
  XMM-Newton}. According to Takahashi et al. (2009), the cool phase
and hot phase coexist and dominate in the
central 70$h_{71}^{-1}$ kpc of the Centaurus cluster, exhibiting a
temperature ratio $T_{\rm
  c}/T_{\rm h} \approx 0.46$. These properties are analogous to those
of A1795. In addition, a similar central
excess in the total gravitating mass profile, with a radius of
50$h_{71}^{-1}$ kpc, has also been found in the Centaurus cluster
(Ikebe et al. 1999). Thus, the two clusters are very similar in the
ICM properties, as well as in the total mass distribution.

As shown in \S3.1.4 and Figure 4{\it b}, the cool ($\sim 2$ keV) phase component of A1795
occupies up to only 20\% volume even in the innermost region, and in  
Figure 6 and 7, no apparent separation on the spatial distribution is seen
between the cool phase and hot phase ICM in the cluster 
core. This indicates that the cool phase is substantially mixed 
into the surrounding hot phase, instead of forming any
``cool-phase-only'' core on scale of $\geq 15$$h_{71}^{-1}$ 
kpc. Since such a
2P structure is seen predominately around a cD galaxy, with an
enhanced metallicity (particularly in the cool phase), the phenomenon
is very likely related to the cD galaxy.


\subsection{Bubble Uplifting Model}

One possible formation mechanism of the 2P structure could be an
AGN-driven gas transport. As indicated by numerical simulations, e.g., Churazov et al. (2001) and Guo
\& Mathews (2010), the buoyant bubbles created by AGN outbursts can drag a certain amount 
of surrounding cool, metal-rich gas to larger radii. For example, the
ICM in central 30$h_{71}^{-1}$ kpc regions of the Hydra A cluster and
M87 are known to be in a 2P or multi-phase form (Ikebe et al. 1997; Molendi
2002). More recently, the distributions of cool phase ICM were
found to coincide with the powerful radio lobes and X-ray cavities, hence, the bubble
uplifting model has successfully
been invoked for both Hydra A cluster and M87 (Nulsen et al. 2002;
Simionescu et al. 2008). These properties may be useful to explain the origin
of the co-existing 2P ICM. As for A1795, by adopting the obtained
distributions of total gravitating mass and cool phase gas
mass ($M(R)$ and $M_{\rm c, gas}$, respectively; \S
3.3.2), the energy required to uplift the cool
component from cD galaxy center to current position (average $R
\approx 40$$h_{71}^{-1}$ kpc) is estimated as $\int_{0}^{R} G M_{\rm c, gas} M(R^{\prime})/{R^{\prime}}^{2} {\rm d}R^{\prime} \sim 10^{58}$ ergs, which
is consistent 
with the amount of mechanical energy injected from its central AGN outburst (Rafferty et al. 2006).
Nevertheless, the faint X-ray cavity detected in A1795 is located
$\lesssim 20$$h_{71}^{-1}$ kpc from the center (Birzan et al. 
2004), apparently much smaller than the radius of the cool phase ICM
as shown in
the obtained 2-D temperature and the $Q_{\rm c}$/$Q_{\rm h}$ maps (Fig. 6
\& 7). Hence the appearance of the 2P ICM cannot be
solely ascribed to bubble uplifting via recent AGN activity. Neither
can the cool phase ICM be bubble remnant, since the uplifted gas would
conductively mix and reach thermal equilibrium with the surrounding
hot phase ICM quickly, i.e., within $\sim 10^{6}$ yrs (e.g., Simionescu  
et al. 2008), after the bubble is torn apart.

\subsection{cD Corona Model}

The most natural and effective way to sustain a stable 2P structure is
to separate the two ICM phases by magnetic fields. In such a case, the two
phases become thermally insulated with each
other, since the gyroradius of a thermal electron in a $1-20$ $\mu$G field
is smaller than its mean free path by about $10-11$ orders of magnitude
(Sarazin 1988). Actually, cluster central regions are often
threaded by rather strong ($\sim 10$ $\mu$G; Ge \& Owen 1993) magnetic
fields, which are possibly related to the cD galaxy. Employing a
general topological classification of such cD-galaxy-related magnetic
field lines into closed and open ones, Makishima et al. (2001)
proposed that the closed loops (with their both ends anchored to the
cD galaxy) are filled with the cool phase ICM, while the open-line
regions, connecting to outer regions, are permeated by the hot
phase. Thus, the cool phase component is expected to emerge as
numerous 
filamentary substructures in the core region, while the hot phase distributes
throughout the vast cluster volume. In addition, the cool phase is
naturally more 
metal-rich than the hot phase in this scenario, as the supernova yield
of the cD galaxy may be largely confined within the loops
(Takahashi et al. 2009), which are surrounded by intruding, less
contaminated hot phase ICM. This ``cD corona'' model is expected to
provide a natural account
for the appearance of metal-rich cool phase ICM in the center of
A1795.

As the cool phase is thermally insulated from the hot phase, and the
radiative transport between the two phases can be ignored (Sarazin
1988), a continuous, efficient heating source located in the central region is required to prevent it
from collapsing due to radiative cooling (see Aschwanden et al. 2001
for a review). As first pointed out by Rosner, Tucker \& Vaiana (1978;
hereafter RTV)
for solar corona, and applied successfully to the Centaurus cluster by
Takahashi et al. (2009), a loop structure as illustrated in Figure 11
has a built-in feedback mechanism to maintain thermal stability for
the cool phase plasma. In order to characterize the loop  
heating mechanism in a quantitative way, we employed an updated
version of RTV model, an analytical
solution to the hydrostatic model introduced in Aschwanden \&
Schrijver (2002; AS02 hereafter) for solar corona and applied it to the 2P
ICM. This model considers a thin, arch-like loop
delineating magnetic field lines, for which the pressure between the
loop-interior and surrounding plasmas are in equilibrium. To constrain the
temperature and density structures of the loop, AS02 assumed mass and
momentum conservations along the loop, as well as a total energy
balance involving radiative loss rate, heating rate,
and thermal conductive flux. The heating rate function has been defined in
form of $E_{\rm H} = E_{\rm 0}$ ${\rm exp} (- h / h_{\rm H})$, where $E_{\rm
  0}$ is heating rate at the bottom of loop, $h$ is
the height in the loop plane, and $h_{\rm H}$ is the scale height of
the heating source (Fig. 11). The analytic solution by AS02 gave a loop
temperature distribution increasing with $h$ in general, and the exact
temperature gradient was predicted to be sensitive to the heating scale
$h_{\rm H}$. As shown in Figure 9 of AS02, for a uniform heating,
i.e., $h_{\rm H} \geq H$, where $H$ is the loop height, the loop temperature
exhibits a mild decrease towards the loop bottom by $>60$\%, while for a small
scale heating at the bottom, i.e., $h_{\rm H} \ll H$, a large
part of the loop becomes near isothermal, with only a sharp, small-scale temperature
drop seen at the footpoints. Such a loop also obey a scaling law, as
defined in Eq.(29) of AS02 (also see Eq.19 below), which describes the
loop maximum temperature $T_{\rm max}$ as a function of $h_{\rm H}$,
$H$, and external
pressure at footpoint $P_{0}$. Another
important property of the loop-confined plasma is that, as shown in
Eq.(30) of AS02, $T_{\rm max}$ shows rather weak dependence on the heating
rate $E_{\rm 0}$, i.e., $T_{\rm max} \propto E_{\rm 0}^{2/7}$,
because a decrease (or increase) in $E_{\rm 0}$ will 
make the loop thinner (or thicker), so as to keep the plasma
luminosity nearly equal to $E_{\rm 0}$.


Specifically, following AS02, the temperature distribution of the
loop-interior plasma is expected as
\begin{equation} \label{eq:ltp}
T(h)=T_{\rm max}\left\{1-\left[1-\left(\frac{2}{\pi}\right){\rm arcsin}\left(\frac{h}{H}\right)\right]^{a(H/h_{\rm H})}\right\}^{b(H/h_{\rm H})},
\end{equation}
and the density distribution can be calculated as

\begin{equation} \label{eq:ldp}
n(h)=\frac{P(h)}{k_{\rm B}T(h)}=\frac{P_{0}}{k_{\rm B}T(h)}{\rm
  exp}\left[-\frac{\mu m_{\rm p}}{2 k_{\rm
      B}}\int_{0}^{h}\frac{g(h^{\prime})}{T(h^{\prime})}d h^{\prime} \right],
\end{equation}
where
\begin{equation} \label{eq:sl1}
T_{\rm max}=\left(\frac{\pi H}{2}P_{0}\right)^{1/3}S_1^{AS}(H,h_{\rm H})
\end{equation}
is directly derived from the scaling law (Eq.29 of AS02), $g$ is the
gravitational 
acceleration calculated from the obtained total mass profile (\S 3.3.2), 
$a(H/h_{\rm H})$ and $b(H/h_{\rm H})$ are the best-fit analytic
functions, as defined in Eqs.(20) and (21) of AS02, respectively, to
reproduce the dependence of $T(h)$ on $H/h_{\rm H}$ derived in a 
numerical way, and similarly,  
$S_1^{AS}(H,h_{\rm H})$ is the best-fit empirical function of scaling law
factor given in Eq.(35) of AS02. We adopted the best-fit coefficients
of the hydrostatic solutions presented in Table 1 of AS02 to
characterize $a(H/h_{\rm H})$, $b(H/h_{\rm H})$, and
$S_1^{AS}(H,h_{\rm H})$, and then, to calculate $T_{\rm max}$,
$T(h)/T_{\rm max}$, and $n(h)$ based on the loop properties.  
In short, given a set of ($P_{0},H,h_{\rm H}$), we may determine the
temperature and density distributions 
of the loop-interior plasma.

Next we applied the AS02 model to the X-ray observations of A1795 and,
for a comparison, the Centaurus cluster, both harboring a prominent
cool phase in the central region. In the
calculation we determined the external pressures $P_{0}$ by the
temperatures and densities of the hot 
phases measured at the innermost regions, and took the loop heights
$H\approx80$$h_{71}^{-1}$ kpc and 70$h_{71}^{-1}$ kpc for A1795 and the 
Centaurus cluster, respectively,
as indicated by the radii of cool phases (\S 3.1.2; Takahashi et
al. 2009). Figure 12{\it a} shows the predictions of Eq.(19), i.e., $T_{\rm max}$ as a
function of $h_{\rm H}$, in comparison with the measured maximum cool
phase temperature, i.e., $T_{\rm max} \approx 2.4$ keV 
and $2.2$ keV for A1795 and the Centaurus cluster (Fig. 4{\it a};
Takahashi et al. 2009), respectively. This comparison yields $h_{\rm H} \approx
11$$h_{71}^{-1}$ kpc for A1795 and 18$h_{71}^{-1}$ kpc for the
Centaurus cluster, which indicates
that the heating is concentrated at the bottoms of the loops.

Since the sizes of radio lobes of the central AGNs in A1795 and the
Centaurus cluster ($\approx 10$$h_{71}^{-1}$ kpc and
$15$$h_{71}^{-1}$ kpc, respectively; Ge \& Owen 1993; Taylor et
al. 2002) agree well with $h_{\rm H}$
estimated above, we speculate that the AGN feedback is a candidate
heating source for the coronal loops. The life time of the radio lobes can be estimated by
$t_{\rm sync} = \frac{9 m_{\rm e}^3 c^5}{4 e^4 \bar{B}^2 \gamma_{\rm
    e}} \sim 10^{7} - 10^{8}$ yrs, roughly consistent with the
  cooling time of the cool phase ICM. Furthermore, as shown in
  Rafferty et al. (2006), the heating rates provided by the central AGN
  outbursts in A1795 and the Centaurus cluster were estimated to be
  $\simeq 1.6 \times
  10^{44}$ ergs $\rm s^{-1}$ and $7.4 \times 10^{42}$ ergs $\rm
  s^{-1}$, respectively, which agree well with their cool phase luminosities
  ($\simeq 1.4\times10^{44}$ ergs $\rm s^{-1}$ for A1795 and $1.0 \times 10^{43}$
  ergs $\rm s^{-1}$ for the Centaurus cluster).

Having obtained the estimates of $H$ and $h_{\rm H}$ for the two
clusters, we then calculated, via Eq.(17), the $T(h)/T_{\rm max}$
profiles, and show the results in Figure 12{\it b}. Thus, a mild
inward temperature decrease is expected, especially for the Centaurus cluster. Actually, Takahashi et
al. (2009) showed (in their Fig. 4{\it a}) that the cool phase
temperature of the Centaurus cluster decreases mildly, from $\sim
2.2$ keV (as already used in Fig. 12{\it a}) at $\sim 25$$h_{71}^{-1}$ kpc 
, to $\simeq 1.6$ keV = 0.72 $T_{\rm max}$ at $\lesssim
7$$h_{71}^{-1}$ kpc. This latter value, when plotted on Figure 12{\it
  b} (with a dashed cross), agrees with the prediction. The same is
true in the A1795 case, which shows a more flat $T(h)/T_{\rm max}$
profile that is consistent with the measured value of $\simeq 0.94$ at
15$h_{71}^{-1}$ kpc (solid cross on Fig. 12{\it b}). As shown in
Figure 12{\it c}, we applied the obtained $T(h)$ to Eq.(18) and
calculated the normalized loop gas density profile $n(h)/n_{\rm 0}$ for
A1795, which again agrees well with the observed profile.

In summary, the cD corona view, combined with the AS02 modeling of the
loop-confined plasmas, can account for several important properties of
inner regions of A1795 and the Centaurus cluster. These properties
include the stable 2P structure, the higher metallicity of
the cool phase, the absolute values of $T_{\rm c}$, and the spatial
$T_{\rm c}$ and density distributions.  


\section{CONCLUSION}

By analyzing the {\em Chandra}, {\em XMM-Newton}, and
{\em Suzaku} data of the X-ray bright galaxy cluster A1795, we report
clear preference for the 2P ICM model in the central 80$h_{71}^{-1}$
kpc, which consists of a cool phase 
($2.0-2.2$ keV) and a hot phase ($5.0-5.7$ keV) component. This 2P
model provides significantly better fit to the deprojected spectra 
than the 1P model with continuous temperature profile, while the
latter cannot be fully ruled out based on current data. Combining the
{\em Suzaku} XIS and the {\em XMM-Newton} EPIC \& RGS, we 
have marginally detected a third weak 0.8 keV component in the inner
144$h_{71}^{-1}$ kpc region that can be ascribed to a portion of ISM
component of the cD galaxy. Based on a 2-D spectral
analysis with the ACIS data, we have revealed a 
possible spatial correlation between the cool 
phase and metal-rich gas in the $50-100$$h_{71}^{-1}$ kpc
region. A follow-up XIS analysis shows consistent result, that the
cool phase ICM does exhibit a higher metallicity ($\approx 0.80$ 
$Z_\odot$) than the hot phase one ($\approx 0.36$ $Z_\odot$). Hence, we have 
successfully resolved a 2 keV, metal-rich component associated with the 
cD galaxy, which is spatially mixed but thermally separated with 
the surrounding 5 keV cluster component.   
All these properties can be explained by the cD corona model incorporating with 
the AS02 solution for quiescent coronal loops.

\section*{Acknowledgments}
This work was supported by the Grant-in-Aid for Scientific Research (S),
No. 18104004, titled "Study of Interactions between Galaxies and
Intra-Cluster Plasmas", and by the National Science Foundation of China 
(Grant No. 10878001 and 10973010, and National Science Fund for 
Distinguished Young  Scholars) and the Ministry of Science and Technology 
of China (Grant No. 2009CB824900 and 2009CB824904). L. G. was
supported by the Grand-in-Aid for JSPS fellows,  
through the JSPS Postdoctoral Fellowship program for Foreign Researchers.


\begin{deluxetable}{lcccccc}
\tabletypesize{\scriptsize} \tablecaption{Summary of {\it Chandra},
{\it XMM-Newton}, and {\it Suzaku} Observations
\label{tbl:ObsLog}} \tablewidth{0pt} \tablecolumns{10} \tablehead{
\colhead{Date} & \colhead{Detector} & \colhead{ObsID} &
\colhead{RA} & \colhead{Dec} &
\colhead{Raw/Clean Exposure}\\
\colhead{dd mm yyyy} & \colhead{} & \colhead{}
& \colhead{(h m s; J2000)} & \colhead{(d m s; J2000)} & \colhead{(ks)} } \
\startdata

20/12/1999 & {\it Chandra} ACIS-S & 494 & 13 48 56.5 & +26 36 26.1 & 19.8/17.5 \\
21/03/2000 & {\it Chandra} ACIS-S & 493 & 13 48 49.2 & +26 36 27.3 & 19.9/18.9 \\
10/06/2002 & {\it Chandra} ACIS-S & 3666 & 13 48 48.9 & +26 34 32.3 & 14.6/13.7 \\
14/01/2004 & {\it Chandra} ACIS-S & 5287 & 13 48 55.0 & +26 36 35.0 & 14.5/13.5 \\
18/01/2004 & {\it Chandra} ACIS-I & 5289 & 13 48 55.1 & +26 36 45.0 & 15.2/14.7 \\
23/01/2004 & {\it Chandra} ACIS-I & 5290 & 13 49 00.8 & +26 42 07.5 & 15.1/14.7 \\
20/03/2005 & {\it Chandra} ACIS-I & 6159 & 13 48 32.9 & +26 40 45.4 & 15.1/14.6 \\
20/03/2005 & {\it Chandra} ACIS-S & 6160 & 13 48 49.4 & +26 36 27.1 & 15.0/14.1 \\
28/03/2005 & {\it Chandra} ACIS-I & 6161 & 13 49 19.5 & +26 31 05.4 & 13.8/13.3 \\
28/03/2005 & {\it Chandra} ACIS-I & 6162 & 13 48 48.0 & +26 36 21.7 & 13.8/13.3 \\
31/03/2005 & {\it Chandra} ACIS-I & 6163 & 13 48 47.6 & +26 36 16.4 & 15.1/14.8 \\
26/06/2000 & {\it XMM-Newton} EPIC-pn & 0097820101 & 13 48 52.3 & +26 35 21.6 & 47.7/30.2 \\
26/06/2000 & {\it XMM-Newton} EPIC-MOS1 & 0097820101 & 13 48 52.3 & +26 35 21.6 & 50.1/43.6\\
26/06/2000 & {\it XMM-Newton} EPIC-MOS2 & 0097820101 & 13 48 52.3 & +26 35 21.6 & 50.1/40.3 \\
26/06/2000 & {\it XMM-Newton} RGS & 0097820101 & 13 48 52.3 & +26 35 21.6 & 66.6/40.1 \\
13/01/2003 & {\it XMM-Newton} EPIC-pn & 0109070201 & 13 48 40.0 & +26 22 16.4 & 70.3/48.1 \\
13/01/2003 & {\it XMM-Newton} EPIC-MOS1 & 0109070201 & 13 48 40.0 & +26 22 16.4 & 64.4/54.1 \\
13/01/2003 & {\it XMM-Newton} EPIC-MOS2 & 0109070201 & 13 48 40.0 & +26 22 16.4 & 64.4/54.2 \\
10/12/2005 & {\it Suzaku} XIS & 800012010 & 13 48 53.8 & +26 36 03.6 & 13.1/13.0 \\
10/12/2005 & {\it Suzaku} XIS & 800012020 & 13 48 53.3 & +26 47 57.5 & 24.4/24.0 \\
11/12/2005 & {\it Suzaku} XIS & 800012030 & 13 48 53.5 & +26 59 58.2 & 30.6/$--$\tablenotemark{a} \\
11/12/2005 & {\it Suzaku} XIS & 800012040 & 13 48 53.5 & +26 24 02.5 & 26.1/25.5 \\
12/12/2005 & {\it Suzaku} XIS & 800012050 & 13 48 53.5 & +26 12 00.4 & 40.1/37.2 \\
10/12/2005 & {\it Suzaku} HXD & 800012010 & 13 48 53.8 & +26 36 03.6 & 10.4/10.4 \\

\enddata
\tablenotetext{a}{This dataset is not used in analysis. See
  text \S2.1.3.}
\end{deluxetable}

\begin{deluxetable}{cccccccccc}

\centering \tabletypesize{\scriptsize} \tablewidth{0pt}
 \tablecaption{Best-fit Temperature and Metal Abundance Gradients
   Obtained with 1P/2P APEC Fittings\tablenotemark{a}}

 \tablehead{         Region &  $f_{\rm ICM}$ & $T_{1P}$ & $A_{1P}$ &
   $\chi_{1P}^2/\nu_{1P}$ & $T_{c}$ & $T_{h}$ & $A_{2P}$ &
   $\chi_{2P}^2/\nu_{2P}$  \\
   ($h_{71}^{-1}$ kpc) & & (keV) & ($Z_\odot$) & & (keV) & (keV) &
   ($Z_\odot$) 
}

\startdata
& \multicolumn{7}{c}{{\it Chandra} ACIS} & \\
\cline{2-6}
$0-30$ & 99.8\% & $3.27^{+0.27}_{-0.31}$ & $0.64^{+0.10}_{-0.09}$ & 1750/1496 &
$2.18^{+0.45}_{-0.18}$ & $6.28^{+1.86}_{-1.80}$ & 
$0.54^{+0.11}_{-0.12}$ & 1608/1494 \\
$30-51$ & 99.6\% & $3.81^{+0.27}_{-0.22}$ & $0.84^{+0.10}_{-0.09}$ &--- &
$2.51^{+0.55}_{-0.31}$ & $5.88^{+0.92}_{-1.00}$ &
$0.74^{+0.13}_{-0.12}$ & --- \\
$51-80$ & 99.0\% & $4.67^{+0.26}_{-0.27}$ & $0.60^{+0.09}_{-0.10}$ &--- &
$2.12^{+0.40}_{-0.44}$ & $5.99^{+0.64}_{-0.84}$ &
$0.51^{+0.09}_{-0.11}$ & --- \\
$80-116$ & 97.8\% & $5.61^{+0.37}_{-0.38}$ & $0.56^{+0.08}_{-0.08}$ &--- &
--- & ---& ---& --- \\
$116-238$ & 93.4\% & $6.52^{+0.32}_{-0.31}$ & $0.36^{+0.11}_{-0.10}$ &--- &
--- & ---& ---& ---\\
$238-354$ & 86.7\% & $6.19^{+0.34}_{-0.39}$ & $0.35^{+0.11}_{-0.10}$ &--- &
--- & ---& ---& ---\\
$354-707$ & 82.4\% & $5.31^{+0.52}_{-0.46}$ & $0.28^{+0.06}_{-0.07}$ &--- &
--- & ---& ---& ---\\
$707-1335$ & 10.2\% & $3.42^{+0.77}_{-0.62}$ & $0.23^{+0.12}_{-0.12}$&--- &
--- & ---& ---& --- \\
\hline
& \multicolumn{7}{c}{{\it XMM-Newton} EPIC} & \\
\cline{3-7}
$0-30$ & 99.8\% & $3.07^{+0.07}_{-0.07}$ & $0.52^{+0.06}_{-0.05}$ & 1040/881 &
$2.04^{+0.53}_{-0.64}$ & $4.78^{+2.54}_{-1.04}$ &
$0.45^{+0.06}_{-0.06}$ & 929/879 \\
$30-51$ & 99.7\% & $3.67^{+0.06}_{-0.06}$ & $0.61^{+0.06}_{-0.06}$ & --- &
$2.00^{+0.41}_{-0.55}$ & $4.70^{+0.65}_{-0.61}$ & 
$0.58^{+0.06}_{-0.05}$ & --- \\
$51-80$ & 99.5\% &  $4.16^{+0.08}_{-0.08}$ & $0.50^{+0.07}_{-0.05}$ & --- &
$1.83^{+0.53}_{-0.38}$ & $5.10^{+0.56}_{-0.30}$ & 
$0.51^{+0.05}_{-0.05}$ & --- \\
$80-116$ & 99.1\% & $5.03^{+0.12}_{-0.12}$ & $0.50^{+0.04}_{-0.05}$ &--- &
--- & ---& ---& ---\\
$116-238$ & 97.3\% & $5.32^{+0.10}_{-0.10}$ & $0.40^{+0.06}_{-0.07}$ &--- &
--- & ---& ---& ---\\
$238-354$ & 91.7\% & $5.78^{+0.19}_{-0.19}$ & $0.27^{+0.06}_{-0.05}$ &--- &
--- & ---& ---& ---\\
$354-707$ & 85.5\% & $5.21^{+0.32}_{-0.36}$ & $0.25^{+0.05}_{-0.06}$ &--- &
--- & ---& ---& ---\\
$707-1335$ & 13.8\% & $2.92^{+0.45}_{-0.52}$ & $0.20^{+0.10}_{-0.09}$ &--- &
--- & ---& ---& ---\\
\hline
& \multicolumn{7}{c}{{\it Suzaku} XIS} & \\
\cline{3-7}
$0-144$ & 99.4\% & $4.69^{+0.06}_{-0.06}$ & $0.48^{+0.03}_{-0.03}$ & 845/775 &
$2.09^{+0.49}_{-0.45}$ & $5.48^{+0.41}_{-0.36}$ &
$0.47^{+0.03}_{-0.03}$ & 802/773 \\
$0-144$\tablenotemark{b} & 99.4\% & --- &--- &--- &
$2.34^{+0.55}_{-0.84}$ & $5.58^{+0.51}_{-0.46}$ & 
$0.48^{+0.03}_{-0.03}$ & 791/771 \\
$144-320$ & 96.4\%  & $5.29^{+0.10}_{-0.10}$ & $0.36^{+0.04}_{-0.03}$ & --- &
[$2.09^{+0.49}_{-0.45}$]\tablenotemark{c} & [$5.48^{+0.41}_{-0.36}$]\tablenotemark{c} &$0.35^{+0.03}_{-0.03}$ & ---\\
$320-700$ & 86.9\% & $5.39^{+0.38}_{-0.37}$ & $0.28^{+0.05}_{-0.04}$ &--- &
--- & ---& ---& ---\\
$700-1444$ & 14.3\% &$3.20^{+0.44}_{-0.43}$ & $0.23^{+0.12}_{-0.11}$ &--- &
--- & ---& ---& ---\\
\hline

\enddata
\scriptsize
\tablenotetext{a}{Fraction of ICM emission in the total $0.5-10.0$ keV
  spectrum(col. [2]), and the best-fit gas temperatures and metal abundance
  gradients obtained by applying 1P ICM model (col. [3] \& [4]) and
  2P ICM
  model (col. [6]$-$[8]) to the ACIS, EPIC, and XIS deprojected spectra. The 1P ICM and 2P ICM
  models are defined in \S3.1.1 and \S3.1.2,
  respectively. $\chi^2/\nu$ of the fits are shown in 
  col. [5] \& [9].  }
\tablenotetext{b}{Best-fit 2P gas temperatures and abundances,
  obtained by applying 2P ICM plus 0.8 keV component model (\S3.1.3) to the
  XIS spectra extracted in $0-144$$h_{71}^{-1}$ kpc region.}
\tablenotetext{c}{The 2P temperatures in this region are fixed to the
  best-fit values in $0-144$$h_{71}^{-1}$ kpc region.}

\end{deluxetable}

\begin{deluxetable}{ccccccccccc}
\centering \tabletypesize{\scriptsize} \tablewidth{0pt}
\tablecaption{Best-fit Metal Abundance and
  Absorption Gradients Obtained with the 2P VAPEC Fittings\tablenotemark{a}}
\tablehead{ \colhead{Region}           & \colhead{O}      &
\colhead{Mg}         & \colhead{Si}     &
\colhead{Fe} & \colhead{$N_{\rm{H, Suzaku}}$} & \colhead{$\chi^2/\nu$}  &
\colhead{$N_{\rm{H, Chandra}}$} & \colhead{$N_{\rm{H, XMM}}$} \\
\colhead{($h_{71}^{-1}$ kpc)} & \colhead{($Z_\odot$)} & \colhead{($Z_\odot$)} &
\colhead{($Z_\odot$)} & \colhead{($Z_\odot$)} & \colhead{$(10^{20}$
  cm$^{-2})$} &\colhead{} & \colhead{$(10^{20}$ cm$^{-2})$} & \colhead{$(10^{20}$ cm$^{-2})$} 

   }
\startdata

$0-144$   & 0.44$^{+0.23}_{-0.22}$  &
0.53$^{+0.25}_{-0.24}$ & 0.54$^{+0.15}_{-0.15}$ &
0.48$^{+0.03}_{-0.03}$ & 3.26$^{+0.54}_{-0.56}$ 
 & 343/328 & 2.17$^{+0.51}_{-0.49}$ & 1.08$^{+0.33}_{-0.33}$\\

$144-320$ & 0.46$^{+0.24}_{-0.23}$  &
0.50$^{+0.27}_{-0.26}$ & 0.37$^{+0.19}_{-0.19}$ &
0.34$^{+0.03}_{-0.03}$ & 2.07$^{+0.81}_{-0.82}$  &  299/284 & 
1.20$^{+0.24}_{-0.32}$ & 1.34$^{+0.26}_{-0.26}$ \\

$320-700$  & 0.54$^{+0.44}_{-0.42}$  &
0.39$^{+0.52}_{-0.39}$ & 0.10$^{+0.25}_{-0.10}$ &
0.31$^{+0.05}_{-0.05}$  & 1.41$^{+0.51}_{-0.59}$  & 145/149 &
1.21$^{+0.25}_{-0.25}$ & 1.07$^{+0.25}_{-0.25}$ \\

\enddata
\tablenotetext{a}{Best-fit metal abundance (col. [2]$-$col. [5]) and
  absorption gradients (col. [6]), obtained by applying the 2P ICM model to the
  deprojected \suzaku\ XIS spectra extracted in $0-144$$h_{71}^{-1}$ kpc and
  $144-320$$h_{71}^{-1}$ kpc, and by applying the 1P model to the XIS spectra in
  $320-700$$h_{71}^{-1}$ kpc (see text \S3.1.6). $\chi^2/\nu$ of the fits are shown in
  col. [7]. As a comparison, in col. [8] \& [9] we show best-fit absorptions obtained with the
  deprojected \chandra\ ACIS and \xmm\ EPIC spectra, respectively. } 

\end{deluxetable}

\begin{deluxetable}{cccccccccc@{}}

  \tabletypesize{\tiny} \tablecaption{Analysis of X-ray Surface Brightness Profiles}

 \tablehead{            &       & \multicolumn{6}{c}{Projected $\beta$ and Double-$\beta$ Fits\tablenotemark{a}} \\
\hline
& \multicolumn{3}{c}{$\beta$ Model} & & &
\multicolumn{3}{c}{Double-$\beta$ Model} \\
\cline{2-4}
\cline{6-10}
    Energy (keV) &  $R_{\rm c}$\tablenotemark{b} &
    $\beta$&$\chi^2/\nu$& & $R_{\rm c1}$ &$\beta _{\rm 1}$&$R_{\rm c2}$&$\beta_{\rm 2}$&$\chi^2/\nu$}
\startdata
  $0.5-3.0$ & $48.4 \pm 0.9$ & $0.56 \pm 0.01$ & 810/143 & & $53.2 \pm 2.1$ & $0.72 \pm 0.03$
                            & $203.0 \pm 7.2$ & $0.73 \pm 0.05$ & 143/140 \\

   $3.0-8.0$ & $60.6 \pm 1.2$ & $0.60 \pm 0.01$ & 484/121&  & $53.5 \pm 2.6$ & $0.76 \pm 0.04$
                            & $185.5 \pm 16.5$ & $0.81 \pm 0.08$ & 135/118 \\
\hline
\hline
      &       & \multicolumn{6}{c}{Deprojected 1P $\beta$, 1P NFW, and 2P
        $\beta$+Double$-\beta$ Fits\tablenotemark{c}} & & \\
\hline
 & & \multicolumn{3}{c}{1P $\beta$ Model\tablenotemark{d}} & & \multicolumn{3}{c}{1P NFW Model\tablenotemark{d} }  \\

\cline{3-5}
\cline{7-9}
Energy (keV)  & &$R_{\rm c}$&$\beta$&$\chi^2/\nu$& &$R_{\rm s}$ &$c$      &$\chi^2/\nu$\\
\hline
$0.5-3.0$ & & $110.0 \pm 0.4$  & $0.67 \pm 0.01$& 734/143 &  &$281.0 \pm 0.3$&$5.12 \pm 0.10$ &351/143 & \\
$3.0-8.0$ & & $108.0 \pm 0.5$  & $0.70 \pm 0.02$& 364/121 &  &$289.0 \pm 0.3$&$5.12 \pm 0.12$ &286/121 & \\

& & \multicolumn{7}{c}{2P $\beta$+Double$-\beta$ Model}  \\
\cline{3-9}
Energy (keV) & & $R_{\rm c,c}$ &$\beta _{c}$&$R_{\rm c,h1}$ &$\beta _{\rm h1}$&$R_{\rm c,h2}$ &$\beta_{\rm h2}$&$\chi^2/\nu$ &  \\
\hline
$0.5-3.0$ & & $21.2 \pm 4.64$        &$0.60 \pm 0.10$ &$73.1 \pm 2.2$& $0.81 \pm 0.05$ & $274.0 \pm 8.6$            &$0.81 \pm 0.05$ & 146/137 & \\
$3.0-8.0$ & & $[25.1]$\tablenotemark{e}         & $[0.58]$\tablenotemark{e} &$68.3 \pm 3.1$&$0.84 \pm 0.06$ & $211.5 \pm 8.4$            &$0.84 \pm 0.06$ & 143/115 & \\
$0.5-8.0$ & & $25.1 \pm 3.32$         & $0.58 \pm 0.09$ &$72.3 \pm 1.9$&$0.81 \pm 0.05$ & $275.5 \pm 8.7$            &$0.83 \pm 0.06$ & 161/151& \\

 \enddata
\scriptsize
\tablenotetext{a}{Models and corresponding parameters are defined in \S 3.3.1.}
\tablenotetext{b}{Best-fit $R_{\rm c}$/$R_{\rm c,1}$/$R_{\rm c,2}$/$R_{\rm s}$/$R_{\rm c,c}$/$R_{\rm c,h1}$/$R_{\rm c,h2}$ are given in the unit of $h_{71}^{-1}$ kpc.}
\tablenotetext{c}{Models and corresponding parameters are defined in \S 3.3.1.}
\tablenotetext{d}{Surface brightness profiles were extracted
from $>$ 100$h_{71}^{-1}$ kpc regions.}
\tablenotetext{e}{Due to the insufficient cool phase counts in
  $3.0-8.0$ keV, the cool phase $\beta$ parameters were fixed to the
  best-fit $0.5-8.0$ keV results. }
\end{deluxetable}

\clearpage

\begin{figure}
\begin{center}
\includegraphics[angle=-0,scale=.8]{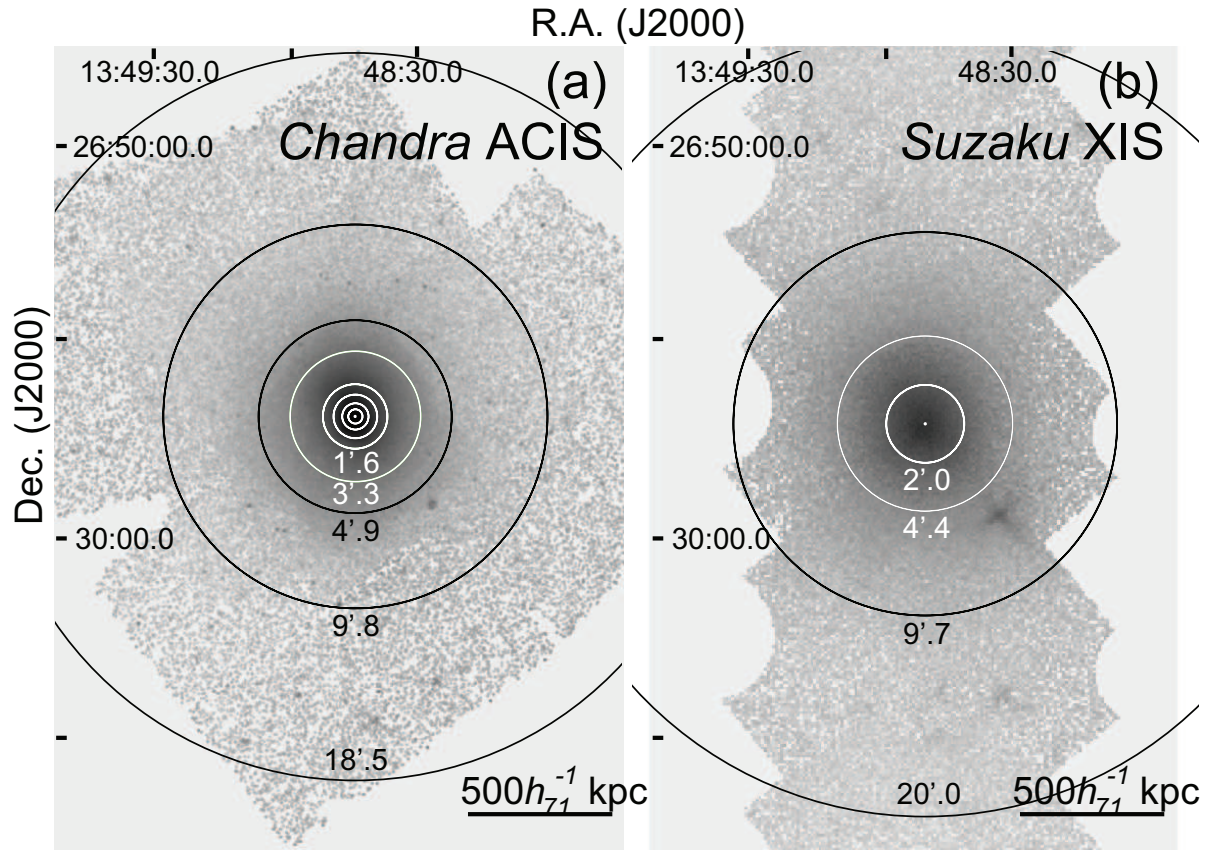}
\caption{Exposure-corrected $0.7-8.0$ keV images of A1795 obtained
  with (a) the \chandra\
ACIS and (b) the \suzaku\ XIS, where the thin and thick annuli used in the
respective spectral analysis are also plotted. 1\arcmin\ corresponds to about 72$h_{71}^{-1}$ kpc.}
\end{center}
\end{figure}

\begin{figure}
\begin{center}
\includegraphics[scale=.3]{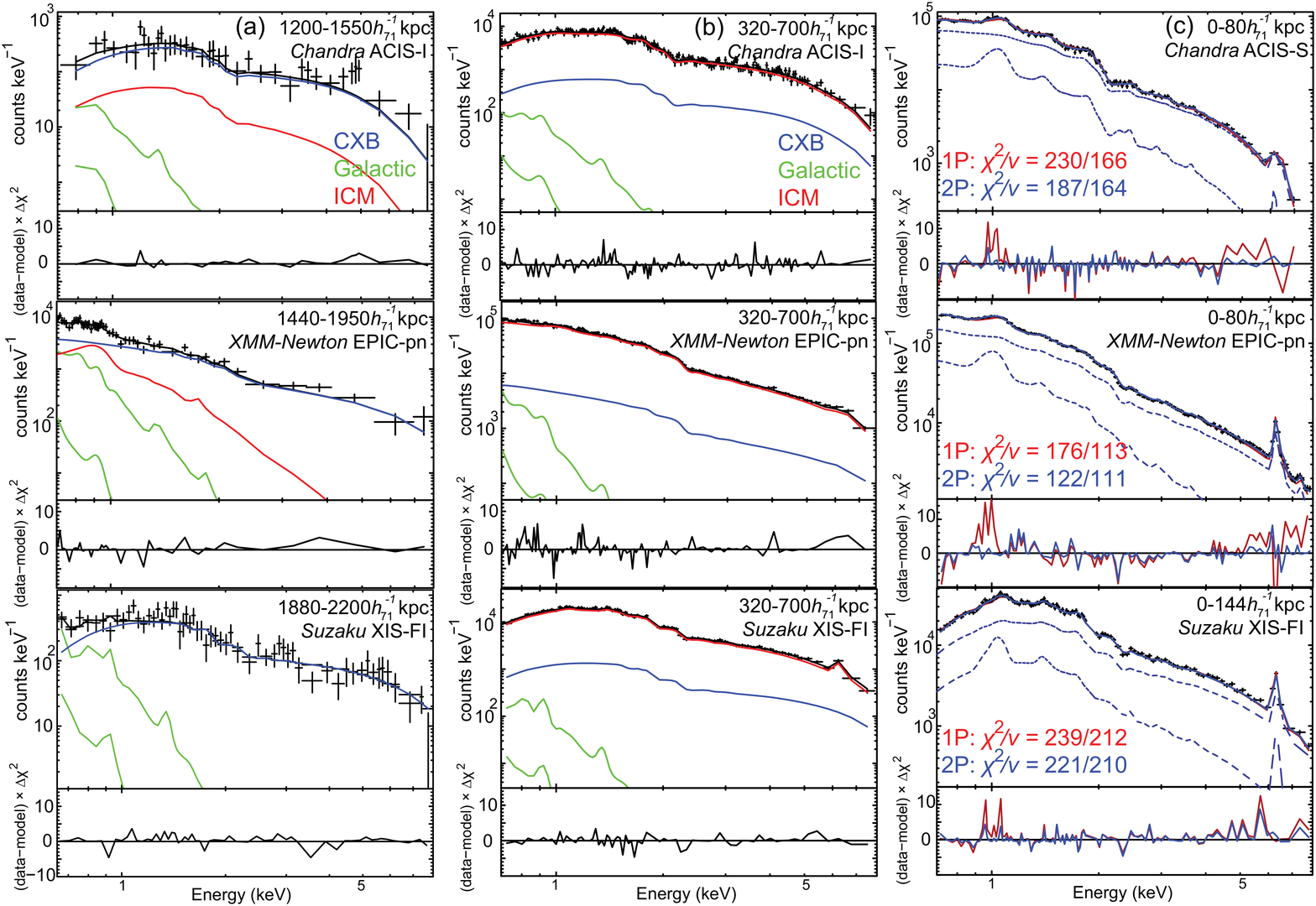}
\caption{(a) NXB-subtracted \chandra\ ACIS-I (upper), \xmm\
EPIC-pn (middle), and \suzaku\ XIS-FI (lower) spectra extracted from
$>1200$$h_{71}^{-1}$ kpc regions that were
used to create the background template (\S 2.2), plotted against the best-fit models
composed of CXB (blue), Galactic (green), and ICM (red) components. (b) NXB-subtracted deprojected \chandra\
ACIS-I (upper),
\xmm\ EPIC-pn (middle), and \suzaku\ XIS-FI (lower) spectra extracted from $320-700$$h_{71}^{-1}$ kpc, along with the
corresponding best-fit CXB (blue), Galactic (green), and ICM (red) models. (c) NXB-subtracted deprojected
\chandra\ ACIS-S (upper), \xmm\ EPIC-pn (middle), and \suzaku\ XIS-FI (lower) spectra
extracted from the central $80$$h_{71}^{-1}$ kpc, $80$$h_{71}^{-1}$ kpc, and $144$$h_{71}^{-1}$ kpc regions, respectively,
fitted with 1P (red solid; \S 3.1.1) and 2P (blue solid; \S 3.1.2) models. The prediction of each best-fit 2P component
is shown with a blue dash line. For clarity the CXB and Galactic
emission models are omitted from the plots.}
\end{center}
\end{figure}

\begin{figure}
\begin{center}
\includegraphics[angle=-0,scale=.35]{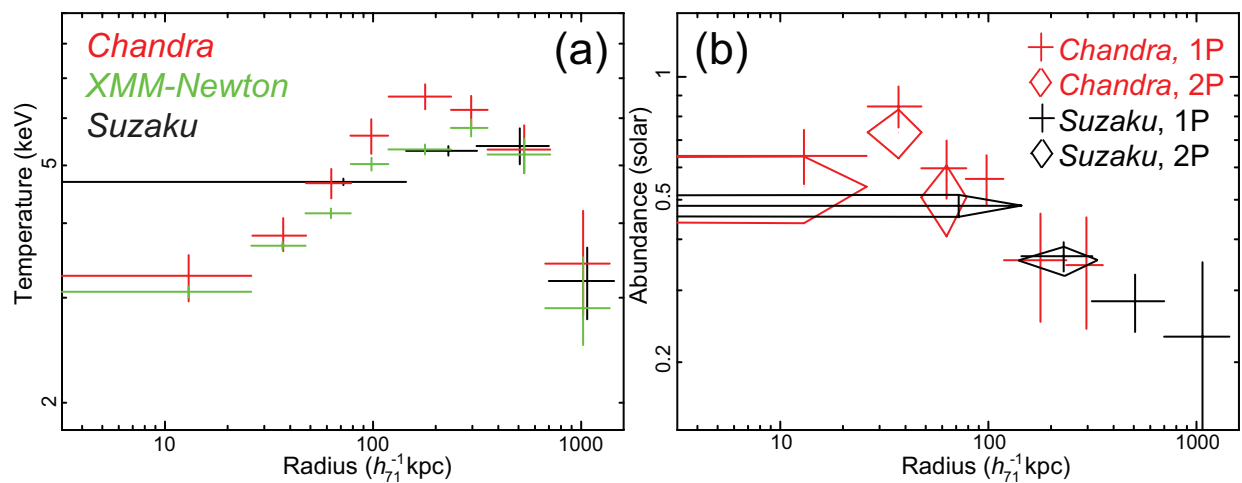}
\caption{Deprojected profiles of (a) gas temperature and
(b) abundance, obtained with the \chandra\ ACIS (red),
\xmm\ EPIC (green), and \suzaku\ XIS (black) data under the
1P ICM assumption (\S 3.1.1). The abundances of the central bins were
also studied with the 2P ICM model (\S 3.1.2), and the results are
plotted in diamonds.}
\end{center}
\end{figure}

\clearpage

\begin{figure}
\begin{center}
\includegraphics[scale=.35]{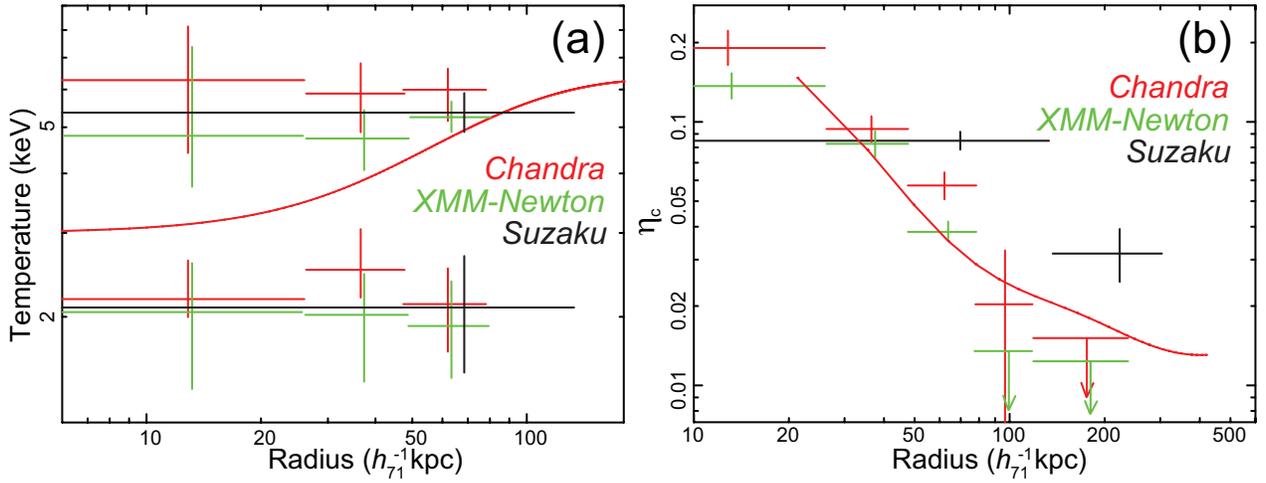}
\caption{(a) Deprojected 2P temperature profiles obtained with the \chandra\ ACIS
(red), \xmm\ EPIC (green), and \suzaku\ XIS (black) data (\S 3.1.2). For comparison,
the 1P result is also plotted with the solid line, which was calculated as the best-fit to the ACIS
1P temperature profile (Fig. 3{\it a}) using the analytic formula in Vikhlinin et al. (2006; Eq.6 therein).
(b) Volume filling factor of the cool component, $\eta_{\rm c}$, 
derived by the deprojected analysis of the \chandra\ ACIS (red),
\xmm\ EPIC (green), and \suzaku\ XIS (black) spectra
(\S 3.1.4). The result obtained
by fitting the ACIS surface brightness in $0.5-8.0$ keV (\S 3.3.1) is plotted
with a solid line. The ACIS and EPIC points on both panels are slightly displaced for
clarity.}
\end{center}
\end{figure}


\begin{figure}
\begin{center}
\includegraphics[angle=-0,scale=.7]{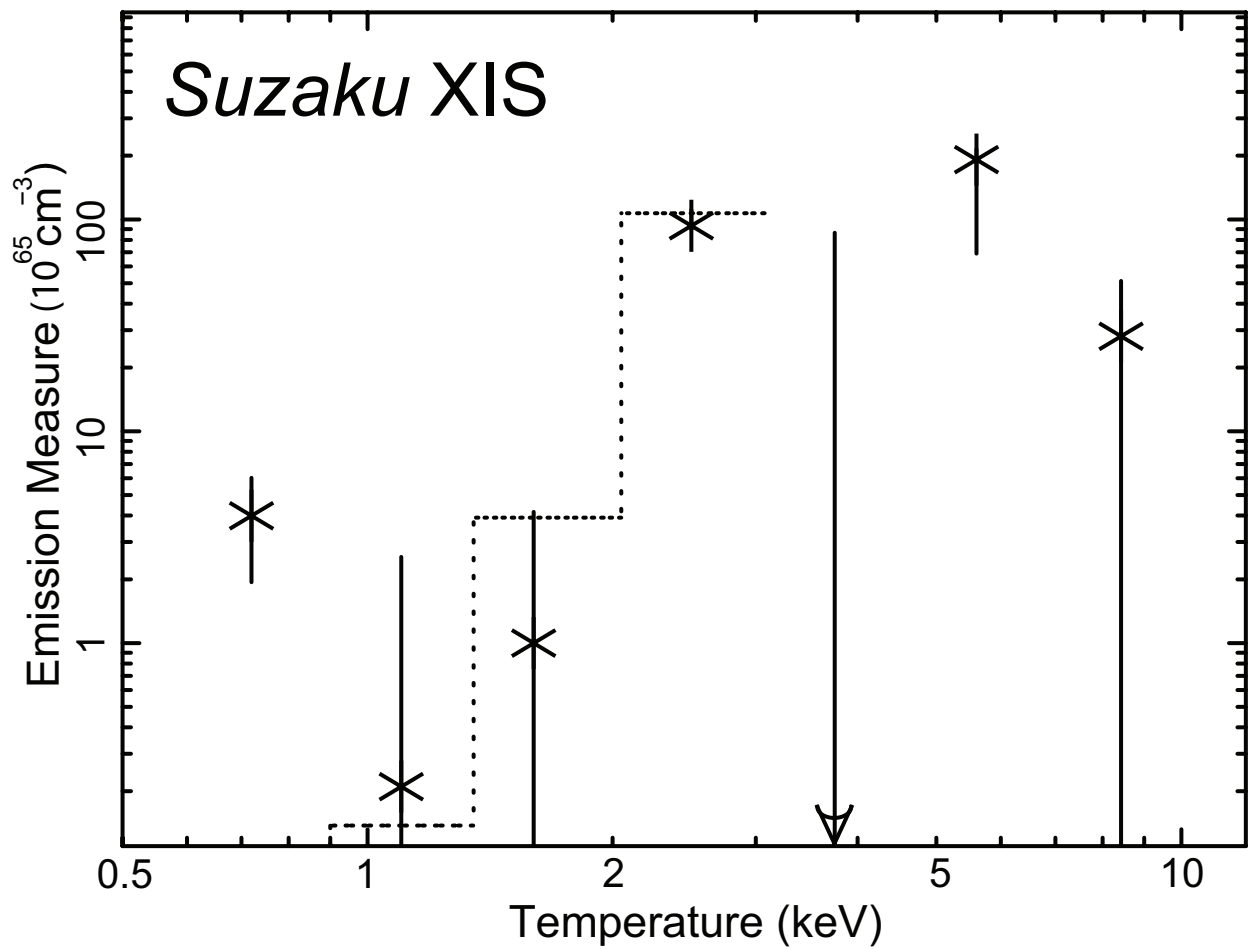}
\caption{Emission measure distribution of different thermal gas
  components in the central 144$h_{71}^{-1}$ kpc region of A1795,
  obtained by fitting a multi-temperature model to the \suzaku\ XIS
  spectra after the projection effect was corrected (\S 3.1.5). Error
  bars were measured at the 68\% confidence
  level. Dotted line shows the prediction of the coronal loop
  model. As described in \S 4.3, this model cannot be used to constrain
  the emission measures of the 0.8 keV component or the hot ICM
  component.}

\end{center}
\end{figure}

\begin{figure}
\begin{center}
\includegraphics[scale=.8]{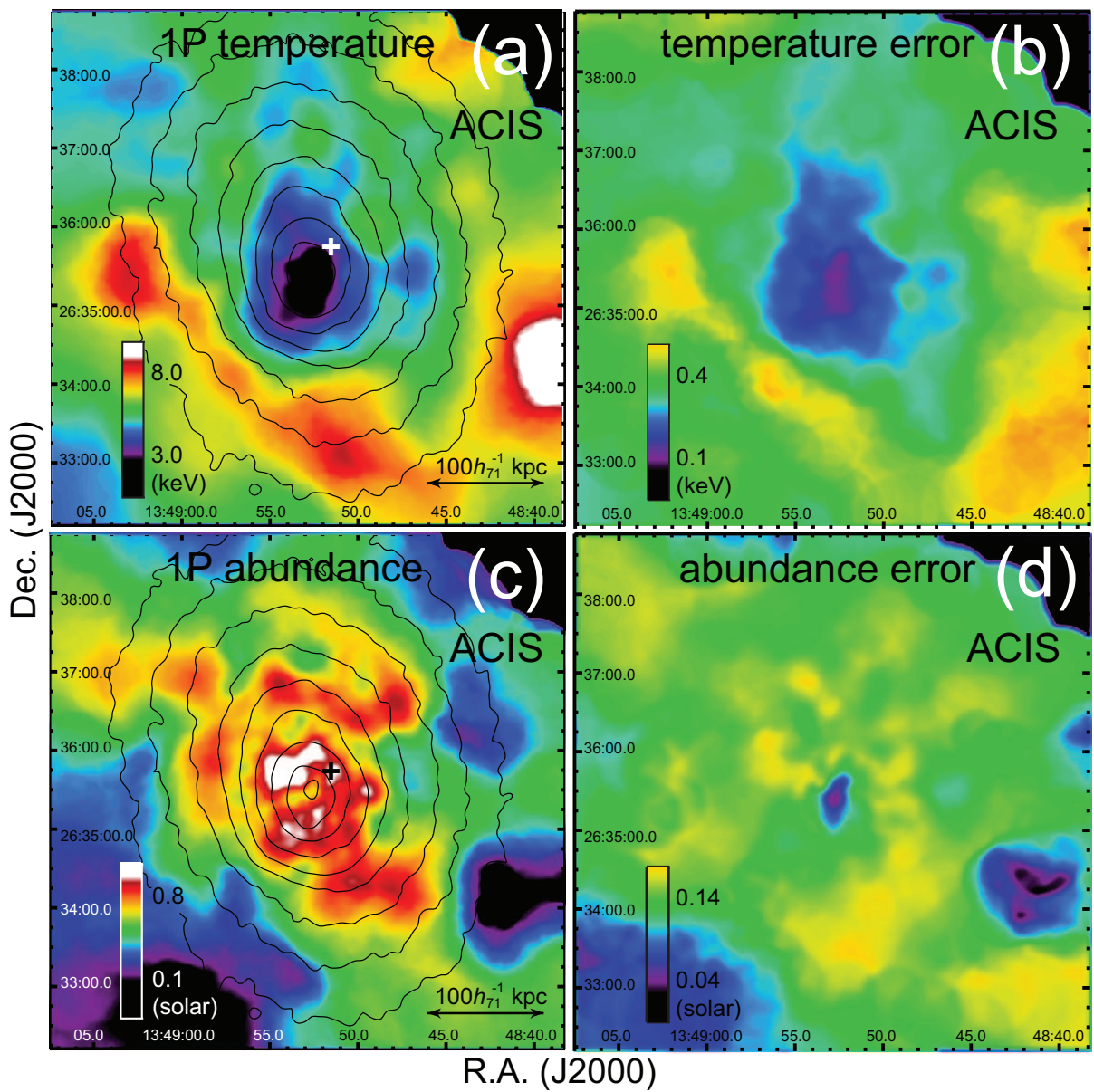}
\caption{(a) A projected gas temperature map, (b) its 68\%
  uncertainties, (c) a gas abundance map, 
and (d) its 68\% uncertainties, all obtained with the 1P spectral analysis of
the ACIS data without incorporating deprojection (\S
3.2.1). In
panels (a) and (c), superposed are the X-ray intensity contours calculated from the $0.5-8.0$ keV data. All maps are centered on the X-ray peak of A1795.
The cross marks the position of the X-ray cavity reported in Birzan et al. (2004).}
\end{center}
\end{figure}
\clearpage

\begin{figure}
\begin{center}
\includegraphics[scale=.8]{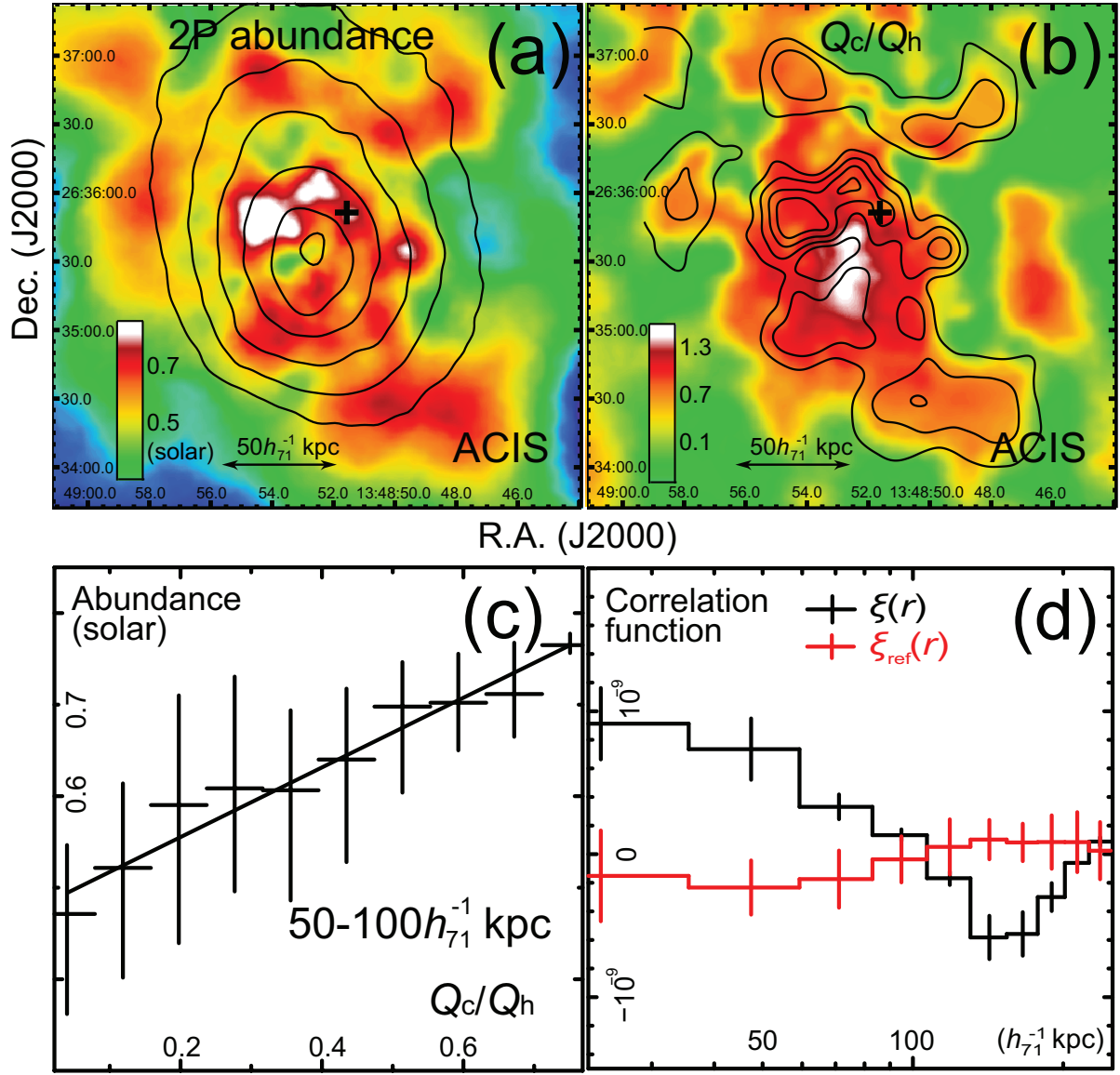}
\caption{A detailed comparison between (a) projected 2P abundance map
and (b) $Q_{\rm c}$/$Q_{\rm h}$ map of the central region, obtained
with the 2P spectral analysis of the
ACIS data without incorporating deprojection (\S 3.2.1). In
panels (a) and (b), the X-ray intensity contours and the 2P
abundance contours are superposed, respectively. The map center and
the meaning of the cross is the same as in Figure 6. (c) A scatter plot between the values in panel (b) and
those in panel (a) in $50-100$$h_{71}^{-1}$ kpc. Error bars are given
at the 68\% confidence level. For clarity, the abundances were
averaged over 10\% intervals in $Q_{\rm c}$/$Q_{\rm h}$. Solid
line shows an analytic fit to the abundance-$Q_{\rm c}$/$Q_{\rm h}$
relation. (d) Two point correlation function between the obtained
2P abundance map and the $Q_{\rm c}$/$Q_{\rm h}$ map (black). A
reference profile (red) is given after randomizing the two maps (see \S 3.2.3 for more details).      
}
\end{center}
\end{figure}

\clearpage

\begin{figure}
\begin{center}
\includegraphics[angle=0,scale=.5]{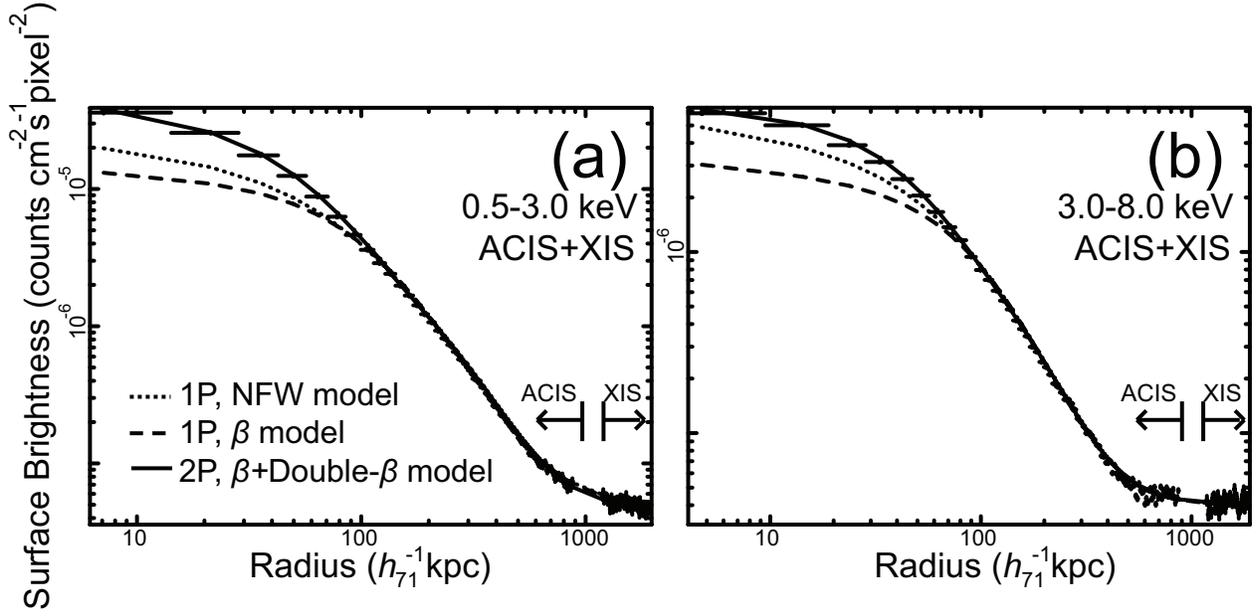}
\caption{Radial surface brightness profiles extracted in (a) $0.5-3.0$ keV 
and (b) $3.0-8.0$ keV,
obtained with the \chandra\ ACIS ($0-1000$$h_{71}^{-1}$ kpc) and \suzaku\
XIS ($1200-1800$$h_{71}^{-1}$ kpc) after the instrumental
normalizations are readjusted (\S 3.3.1).
The best-fit $\beta$+double$-\beta$ models, based on the best-fit deprojected 2P results, are plotted
with solid lines. For comparison, the best-fit $\beta$ model and
  NFW model, based on 1P results that can describe the surface brightness profiles
extracted from
$>$ 100$h_{71}^{-1}$ kpc regions, are also plotted with dash and dotted lines, respectively. }
\end{center}
\end{figure}

\begin{figure}
\begin{center}
\includegraphics[scale=.7]{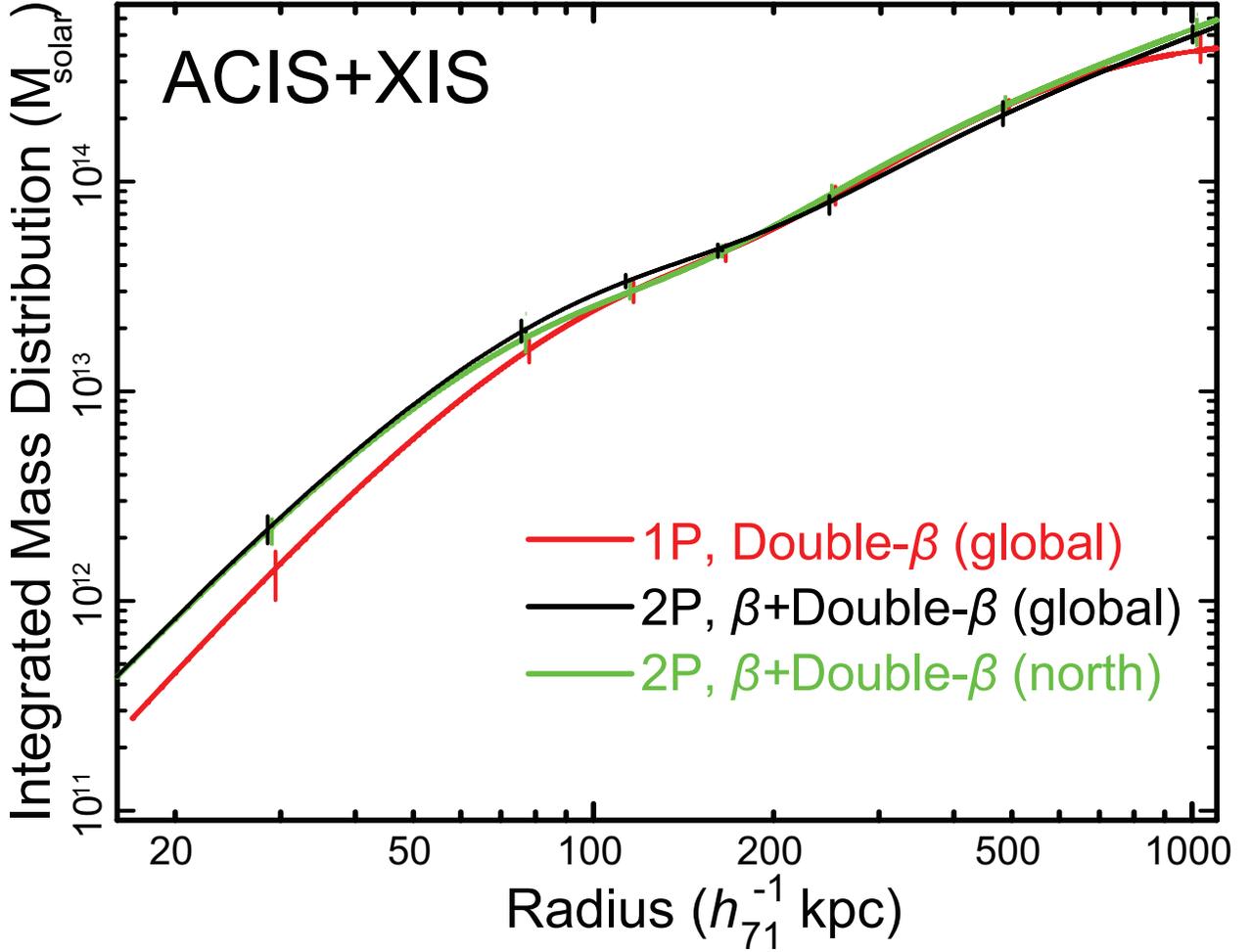}
\caption{Total gravitating mass distributions of A1795, taking into
  account all directions
(black solid line), and northern half only (green solid line), which
were calculated based on the deprojected 2P temperature distribution
and $\beta$+double-$\beta$ 
fitting of the gas density profile (\S 3.1.2 and \S 3.3.1,
respectively). For comparison, the corresponding mass
profiles obtained with the best-fit 1P spectral analysis and
double-$\beta$ gas density models (\S 3.1.1 and \S 3.3.2, respectively)
are also plotted with
red solid lines. Error bars are at the 68\% confidence level.}
\end{center}
\end{figure}

\begin{figure}
\begin{center}
\includegraphics[scale=0.7]{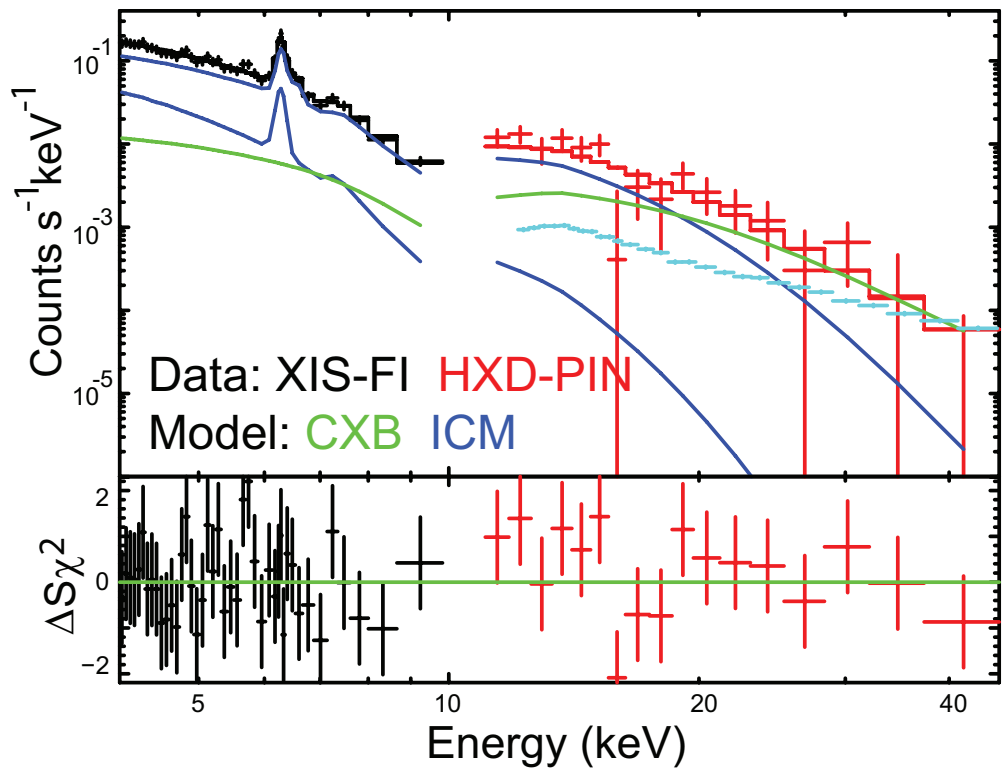}
\caption{NXB-subtracted \suzaku\ XIS-FI spectrum extracted from the central
  320$h_{71}^{-1}$ kpc (black) and HXD-PIN (red) spectrum, compared
with the CXB (green) and ICM (blue) components. The 2P ICM components were
determined in the $<10$ keV range, and extrapolated into the HXD-PIN
range (see \S3.4). The systematic NXB uncertainty of 2.0\% is plotted
in cyan.}
\end{center}
\end{figure}

\begin{figure}
\begin{center}
\includegraphics[scale=.7]{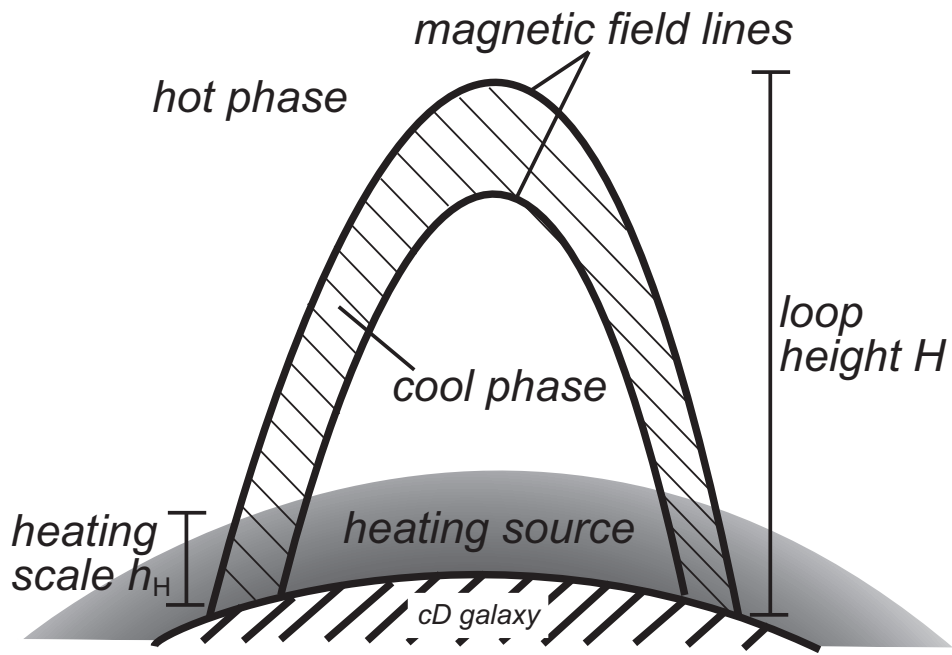}
\caption{A sketch of one of coronal loops that possibly exist in the envelope of the cD galaxy
of A1795 (\S 4.3). $H$ is the loop height and $h_{\rm H}$ is the scale height of the heating source
as shown in gray.}
\end{center}
\end{figure}

\clearpage

\begin{figure}
\begin{center}
\includegraphics[scale=.46]{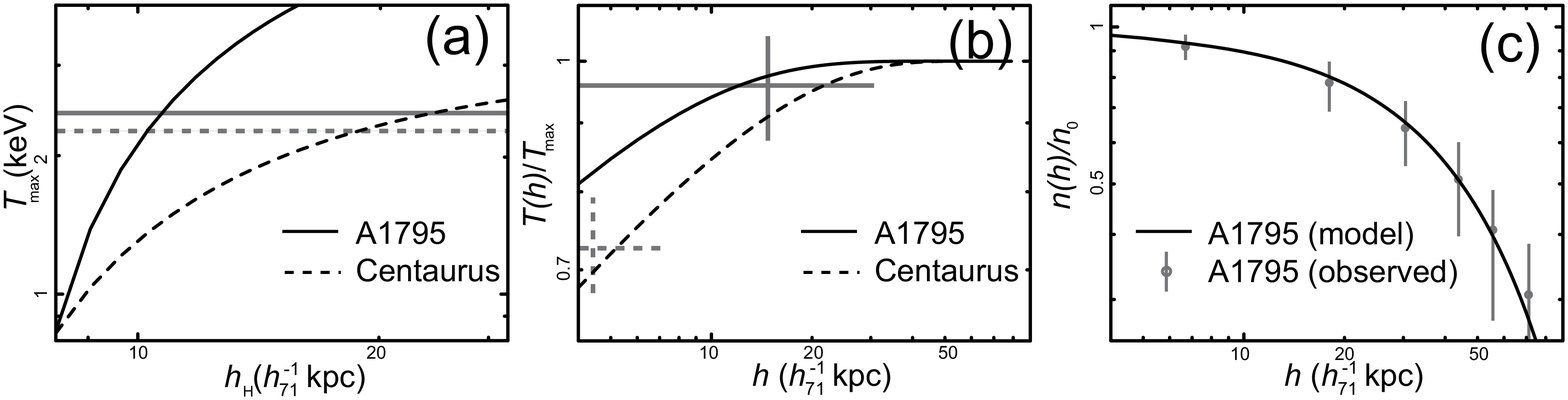}
\caption{(a) Predicted loop maximum temperatures $T_{\rm max}$ as a function of heating scale heights $h_{\rm H}$, calculated
for A1795 (solid) and the Centaurus cluster (dashed). The horizontal lines mark the observed $T_{\rm max}$
for these clusters. (b) Predicted $T(h)/T_{\rm max}$ profiles of the loops for A1795 (solid) and the Centaurus cluster
(dashed), where $h$ is the height in the loop plane. The crosses show the observed $T(h)/T_{\rm max}$ for the innermost shells presented in \S 3.1.2
and Takahashi et al. (2009). (c) Predicted $n(h)/n_{0}$ profile of the loops for A1795 (solid). For
comparison, data points with vertical error bars show the observed $n(h)/n_{0}$ profile 
obtained with the 2P $\beta$+double-$\beta$ fitting of the ACIS surface brightness distribution (\S 3.3.1).}

\end{center}
\end{figure}

\end{document}